\documentclass{article}

\usepackage{arxiv}

\usepackage{booktabs}       % professional-quality tables
\usepackage{amsfonts}       % blackboard math symbols
\usepackage{nicefrac}       % compact symbols for 1/2, etc.
\usepackage{microtype}      % microtypography
\usepackage[utf8]{inputenc} % allow utf-8 input
\usepackage[T1]{fontenc}    % use 8-bit T1 fonts
\usepackage{hyperref}       % hyperlinks
\usepackage{url} 
\usepackage{graphicx}
\usepackage{subcaption}
\usepackage{multirow}
\usepackage{booktabs}
\usepackage{pifont}
\usepackage{enumitem}
\usepackage{makecell}

\usepackage{fancyhdr}

\usepackage{rotating}
\usepackage{array}
\usepackage{lscape}

\newcommand{\cmark}{\ding{51}}
\newcommand{\xmark}{\ding{55}}

\makeatletter
\renewcommand{\@title}{\fontsize{12}{20}\selectfont SoK: Software Debloating Landscape and Future Directions}
\makeatother

\title{\fontsize{15}{20}\selectfont SoK: Software Debloating Landscape and Future Directions}

\author{Mohannad Alhanahnah \\
	Department of Computer Science\\
	University of Wisconsin-Madison\\
	USA \\
	\texttt{mohannad@cs.wisc.edu} \\
	\And
	Yazan Boshmaf \\
	Qatar Computing Research Institute\\
	Hamad Bin Khalifa University\\
	Qatar \\
	\texttt{yboshmaf@hbku.edu.qa} \\
	\And
	Ashish Gehani \\
	Computer Science Laboratory\\
	SRI\\
	USA \\
	\texttt{gehani@csl.sri.com} \\
}

\begin{document}
	\maketitle
	\begin{abstract}
Software debloating seeks to mitigate security risks and improve performance by eliminating unnecessary code. In recent years, a plethora of debloating tools have been developed, creating a dense and varied landscape. Several studies have delved into the literature, focusing on comparative analysis of these tools. To build upon these efforts, this paper presents a comprehensive systematization of knowledge (SoK) of the software debloating landscape. We conceptualize the software debloating workflow, which serves as the basis for developing a multilevel taxonomy. This framework classifies debloating tools according to their input/output artifacts, debloating strategies, and evaluation criteria. 
Lastly, we apply the taxonomy to pinpoint open problems in the field, which, together with the SoK, provide a foundational reference for researchers aiming to improve software security and efficiency through debloating.

\end{abstract}

%\vspace{-0.3cm}
\section{Introduction}
\label{sec:intro}
Modern software development is heavily dependent on third-party libraries to accelerate development and improve functionality~\cite{mohagheghi2004empirical}. However, this practice introduces significant complexity and increases the attack surface of applications due to the integration of various components, each with its own set of dependencies and vulnerabilities~\cite{vlunAnalysis}. The increased complexity increases security risks and leads to code bloat, adversely affecting performance.

Software debloating~\cite{Shankar12,Malecha15}, the process of removing unnecessary code from applications, is a promising approach to address these issues. By eliminating extraneous features, debloating can significantly reduce the attack surface, enhance performance, and improve maintainability. This technique complements other security measures, such as Control-Flow Integrity (CFI)~\cite{ligatti2005control} and Address Space Layout Randomization (ASLR)~\cite{snow2013just}, by minimizing the amount of code that needs protection. Software debloating has gained renewed momentum, in part due to cyber defense initiatives, such as the US Navy's Total Platform Cyber Protection (TPCP) program~\cite{TPCP}. Subsequently, numerous debloating tools were introduced, leading to various studies~\cite{sok_ESORICS,sok_usenix,Hassan23} that examine the literature on software debloating and perform comparative analyses of the prototyped tools. While these studies are thorough, their primary objective is to empirically compare specific aspects, such as resulting binary size or gadget count, of particular types of debloating tools, such as those that target C/C++ programs or containers. The limited scope restricts the influence of these studies to a subset of debloating tools, rather than providing a systematic, comprehensive, and wide-ranging examination of the entire debloating domain, which encompasses a diverse array of tools and evaluation criteria. As such, there is a significant need to augment previous research with a holistic and systematic study of the complete software debloating landscape, thereby enabling more extensive and inclusive conclusions about open issues and challenges in this domain.

To bridge this gap, this paper systematizes the current knowledge on software debloating, providing a multilevel taxonomy that divides the current landscape into three main categories corresponding to the three main stages of the debloating workflow. We also highlight open problems in the field, calling for more practical, usable, and secure debloating solutions that can be integrated seamlessly into modern development workflows.

% Software debloating is a prominent approach that the security community has leveraged to handle the need to reduce the attack surface by (1) identifying unneeded dependencies and (2) automatically removing them. Therefore, there has been a revival of interest in part due to the need for cyber defense (e.g., US Navy Total Platform Cyber Protection (TPCP) program) in extending traditional debloating techniques to reduce code size, improve runtime performance, and remove attack surfaces for a wide spectrum of software applications, native applications, and Docker containers. 
% \input{background}
%\vspace{-0.3cm}
\section{Software Debloating Workflow}
\begin{figure}
  \centering
    \includegraphics[width=0.6\linewidth]{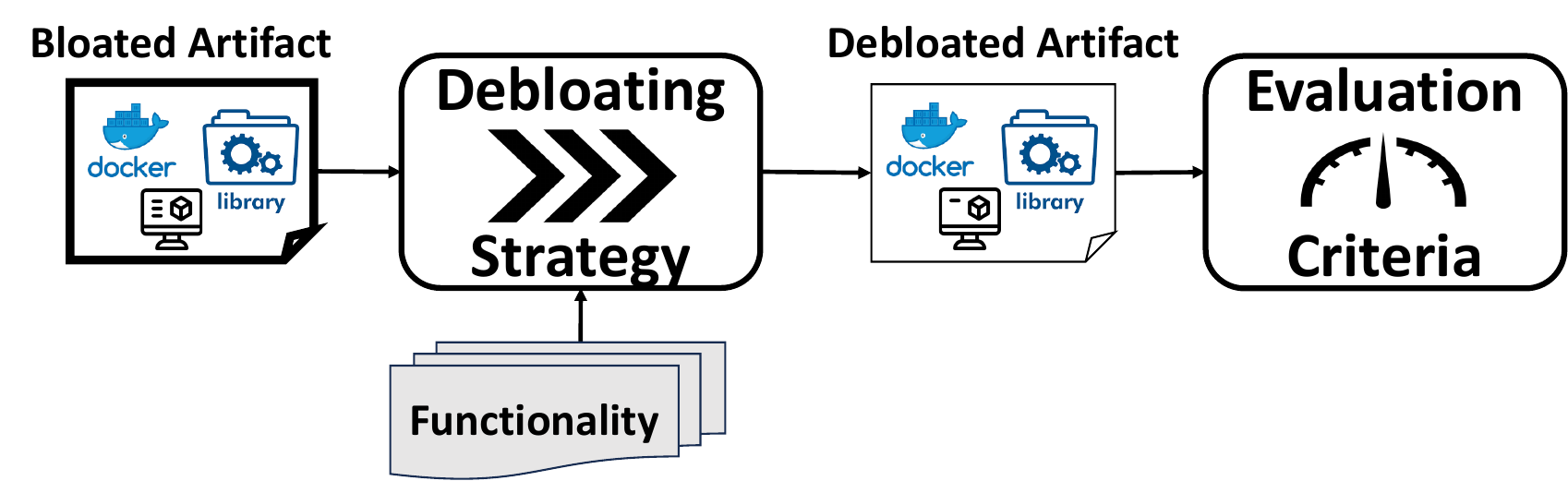}
    %\vspace{-0.1cm}
    \caption{Typical debloating workflow.}
    %\vspace{-0.5cm}
    \label{fig:typicalWorkflow}
\end{figure}

Software bloat refers to unnecessary functionalities and their corresponding software dependencies and components~\cite{Shankar12,Malecha15}. %The debloating process consists of two primary steps: first, identifying these extraneous dependencies, and second, proceeding to eliminate them. 
% however, the debloating process can be extremely challenging, as most programs and libraries do not come with a formal description of their functionality. Consequently, . But these various debloating tools adopted a similar workflow. 
Figure~\ref{fig:typicalWorkflow} depicts the typical workflow used by debloating tools. 
To identify bloat and eliminate it, developers use existing tools that take a \emph{bloated artifact}, such as an application, container, or firmware, often coupled with a \emph{deployment context}, and then produce a \emph{debloated artifact} utilizing a particular \emph{debloating strategy} applied by the tool. After that, the quality of the output artifact is assessed using various evaluation criteria. As depicted in Figure~\ref{fig:typicalWorkflow}, in addition to \emph{bloated artifact}, \emph{debloating strategy} may receive additional input (that is, in the form of annotation or instrumentation) to indicate the required functionality that should be preserved in the output artifact. The next section discusses the details of this debloating workflow in the context of the reviewed literature and our proposed taxonomy.

%\vspace{-0.3cm}

\section{Multi-level Taxonomy} \label{sec:taxonomy}
Our goal is to study and contrast existing software debloating tools and techniques. To achieve this, we first surveyed related research covering all papers published in top-tier security conferences, namely IEEE S\&P, USENIX Security, ACM CCS, and NDSS from 2000 to March 2024. We also selected papers from top academic conferences and journals broadly related to software debloating. This process yielded $48$ publications that are summarized in Table~\ref{table:summary}.

\begin{figure}
  \centering
    \includegraphics[width=\linewidth]{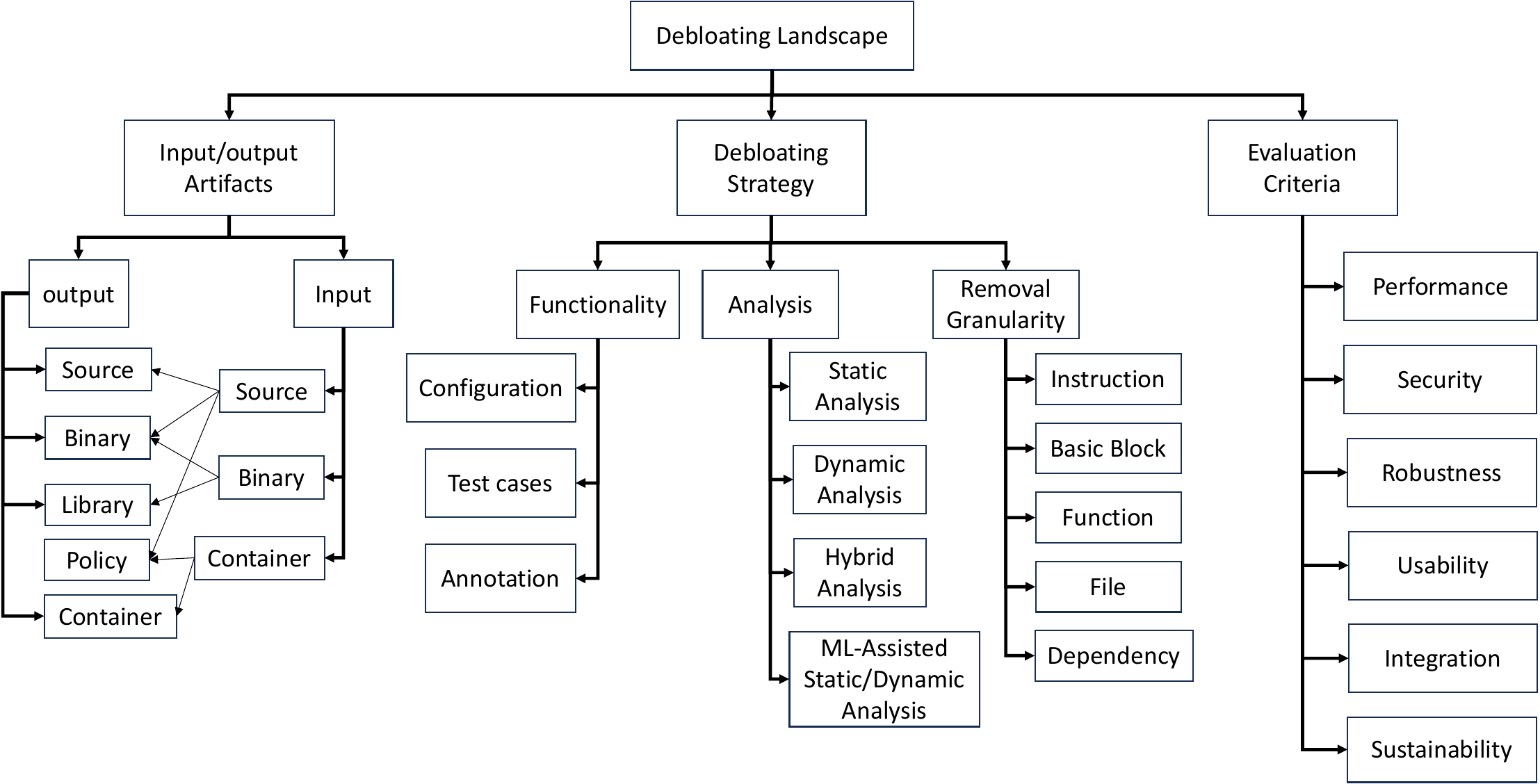}
    %\vspace{-0.6cm}
    \caption{Taxonomy of software debloating landscape.}
    %\vspace{-0.6cm}
    \label{fig:taxonomy}
\end{figure}

Figure~\ref{fig:taxonomy} shows the multilevel taxonomy we designed to categorize the software debloating landscape. 
%We refined the software debloating workflow for this taxonomy, where the evaluation is repeatedly performed and reported back to the debloating strategy to fine-tune or select the best strategy before producing the final output artifact, as depicted in Figure~\ref{fig:proposedDebWF}. 
In this taxonomy, the top level outlines the three main stages of the workflow. Lower levels categorize specific aspects of the debloating landscape, based on the publications listed in Table~\ref{table:summary}, under each stage of the workflow.

%\vspace{-0.2cm}
\subsection{Input/Output Artifacts}

Debloating tools require an input to generate an output. These inputs and outputs are referred to as artifacts and can come in various formats, such as source code, binaries, and containerized applications. The output resulting from debloating can also take any of these forms, or might even be a policy. Figure~\ref{fig:artifacts_count} shows the number of publications with proposed tools that use one or more of the following type mappings between input and output artifacts:

\begin{itemize}
    \item \textbf{Source-to-Source (S2S).} In this workflow, the debloating operation is applied to the given source code, resulting in a minimized source code output. CHISEL~\cite{CHISEL} and Mininode~\cite{Mininode} execute their debloating procedures for C/C++ and JavaScript programs, respectively.  
    \item \textbf{Source-to-Binary (S2B).} The workflow starts with the source code and transforms it into an Intermediate Representation (IR). The debloating process then operates on the IR code. Ultimately, the debloated program is produced in binary format. For instance, LMCAS~\cite{LMCAS} debloats C/C++ programs by first converting them into LLVM IR, resulting in executable output. Tools utilizing this debloating workflow have been applied to platforms such as firmware, as seen with PRAT~\cite{PRAT}. Other tools in this category focus on trimming shared libraries, an approach exemplified by Piece-Wise\cite{Piece-Wise}. Certain tools implementing this workflow extend beyond trimming by incorporating additional checks, such as Saffire~\cite{Saffire}.
    \item \textbf{Artifact-to-Policy (S2P or C2P).} This workflow generates a policy (i.e. \textit{seccomp()}) that limits the program's behavior at run-time. As observed in the reviewed literature, the input artifact for this process can be either source code (S2P) or a containerized application (C2P), as exemplified in debloating tools such as temporal-specialization~\cite{temporal} and Confine~\cite{CONFINE}. Generally, these debloating methods do not involve actual trimming but focus on minimizing the use of unnecessary resources, such as syscalls.
    \item \textbf{Binary-to-Binary (B2B).} The workflow begins with a binary file and results in a debloated program, also in binary format. Similar to S2B tools that apply additional checks, certain tools implementing this workflow extend beyond  trimming, such as Razor~\cite{RAZOR}, and incorporate extra checks, like those of binary control-flow trimming \cite{cf_trimming}, to safeguard CFI. Consequently, the size of the debloated programs may increase in some instances. This debloating approach has been applied to various platforms, including Android (e.g., XDebloat~\cite{XDebloat}, RedDroid~\cite{RedDroid}) and firmware (e.g., IRQDebloat~\cite{IRQDebloat}, DECAF~\cite{DECAF}). A different group of tools focus exclusively on debloating shared libraries, such as BlankIt~\cite{Nibbler} and Nibbler~\cite{Nibbler}. Likewise, tools such as $\mu$Trimmer~\cite{mTrimmer} are designed to debloat shared libraries, but specifically within the context of firmware images.
    \item \textbf{Container-to-Container/s (C2C).} In this workflow, the debloating operation takes a container as input and produces a debloated version of the same container or divides it into multiple containers, each with a portion of the application from the original container. For instance, Cimplifier~\cite{Cimplifier} can function in two modes: either by trimming the container or partitioning it into smaller segments. MMLB~\cite{mmlb} builds on the trimming feature of Cimplifier to investigate bloat in machine learning containers.
    \item \textbf{Dependency-to-Dependency (D2D).}  This workflow accepts inputs consisting of dependency and build management files, like the Project Object Model (POM), where developers outline details about the project, its dependencies, and the build process. The output is a debloated version of the dependency management file(s). An example of this is DepClean~\cite{DEPCLEAN}, which specializes in debloating POM files in Java projects.  
\end{itemize}

\noindent
Some tools adopt a more comprehensive approach to debloat various layers of the software stack, thereby combining multiple type mappings for input/output artifacts. For instance, LightBlue~\cite{LightBlue} debloats the Bluetooth stack, specifically focusing on debloating applications (S2B) and firmware (B2B).

%\vspace{-0.2cm}
\subsection{Debloating Strategies}
This stage of the workflow outlines the methods used by developers to determine unnecessary functionalities, pinpoint their associated dependencies, and remove them. As shown in Figure~\ref{fig:taxonomy}, this stage is divided into three main components, as follows:

\begin{landscape}
	\begin{table*}[h!]
		\centering
		\caption{Selected publications on software debloating landscape.}
		%\vspace{-0.3cm}
		{%\footnotesize
			\scalebox{0.66}{
				\begin{tabular}{|l|l|l|c|c|c|c|c|c|c|c|c|c|c|c|c|c|c|c|}
					\hline
					\multirow{2}{*}{Tool} & \multirow{2}{*}{Venue} & \multirow{2}{*}{\shortstack{In/Out\\ Artifacts}} & \multicolumn{4}{c|}{Removal Granularity} & \multicolumn{3}{c|}{Analysis} & \multicolumn{3}{c|}{Functionality} & \multicolumn{6}{c|}{Evaluation Criteria} \\ \cline{4-19}
					&  &  & \rotatebox{50}{File} & \rotatebox{50}{Library} & \rotatebox{50}{BB} & \rotatebox{50}{Stmt} & \rotatebox{50}{Static} & \rotatebox{50}{Dynamic} & \rotatebox{50}{ML-Assisted} & \rotatebox{50}{Config.} & \rotatebox{50}{Test cases} & \rotatebox{50}{Annotation} & \rotatebox{50}{Performance} & \rotatebox{50}{Security} & \rotatebox{50}{Robustness} & \rotatebox{50}{Usability} & \rotatebox{50}{Integration} & \rotatebox{50}{Sustainability} \\ \hline
					
					Hacksaw~\cite{Hacksaw} & CCS'23 & S2B & \xmark & \cmark & \xmark & \xmark & \cmark & \cmark & \xmark & \xmark & \cmark & \xmark & \xmark & \cmark & \cmark & \xmark & \xmark & \xmark \\
					C2C~\cite{C2C} & CCS'22 & S2P & \xmark & \xmark & \xmark & \xmark & \cmark & \xmark & \xmark & \cmark & \xmark & \cmark & \cmark & \cmark & \xmark & \cmark & \xmark & \xmark \\
					Slimium~\cite{Slimium} & CCS'20 & S2B & \xmark & \cmark & \xmark & \cmark & \cmark & \cmark & \xmark & \xmark & \cmark & \cmark & \xmark & \cmark & \cmark & \cmark & \xmark & \xmark \\
					NA~\cite{cf_trimming} & CCS'19 & B2B & \xmark & \xmark & \cmark & \cmark & \cmark & \cmark & \cmark & \xmark & \cmark & \cmark & \cmark & \cmark & \cmark & \xmark & \xmark & \xmark \\
					CHISEL~\cite{CHISEL} & CCS'18 & S2S & \xmark & \cmark & \cmark & \cmark & \xmark & \cmark & \cmark & \xmark & \cmark & \xmark & \xmark & \cmark & \cmark & \cmark & \xmark & \xmark \\
					PacJam~\cite{PacJam} & ASIACCS '22 & S2S & \cmark & \xmark & \xmark & \cmark & \cmark & \cmark & \xmark & \xmark & \cmark & \xmark & \cmark & \cmark & \xmark & \cmark & \xmark & \xmark \\
					
					LightBlue~\cite{LightBlue} & USENIX Sec'21 & S2B, B2B & \xmark & \cmark & \cmark & \cmark & \cmark & \cmark & \xmark & \cmark & \xmark & \xmark & \xmark & \cmark & \cmark & \xmark & \xmark & \xmark \\
					Temporal Specal.~\cite{temporal} & USENIX Sec'20 & S2P & \xmark & \cmark & \xmark & \xmark & \cmark & \xmark & \xmark & \cmark & \xmark & \xmark & \xmark & \cmark & \xmark & \xmark & \xmark & \xmark \\
					RAZOR~\cite{RAZOR} & USENIX Sec'19 & B2B & \xmark & \xmark & \cmark & \xmark & \cmark & \cmark & \xmark & \xmark & \cmark & \xmark & \cmark & \cmark & \cmark & \cmark & \xmark & \xmark \\ 
					Piece-Wise~\cite{Piece-Wise} & USENIX Sec'18 & S2B & \xmark & \cmark & \xmark & \xmark & \cmark & \cmark & \xmark & \cmark & \xmark & \cmark & \xmark & \cmark & \cmark & \cmark & \xmark & \xmark \\
					
					IRQDebloat~\cite{IRQDebloat} & S\&P'22 & B2B & \xmark & \cmark & \xmark & \xmark & \cmark & \cmark & \xmark & \cmark & \xmark & \cmark & \xmark & \cmark & \xmark & \xmark & \xmark & \xmark \\
					
					LMCAS~\cite{LMCAS} & EuroS\&P'22 & S2B & \xmark & \cmark & \cmark & \cmark & \cmark & \cmark & \xmark & \cmark & \xmark & \cmark & \xmark & \cmark & \cmark & \cmark & \xmark & \xmark \\
					Saffire~\cite{Saffire} & EuroS\&P'20 & S2B & \xmark & \cmark & \xmark & \xmark & \cmark & \cmark & \xmark & \xmark & \xmark & \cmark & \cmark & \cmark & \xmark & \cmark & \xmark & \xmark \\
					
					Mininode~\cite{Mininode} & RAID'20 & S2S & \cmark & \cmark & \xmark & \xmark & \cmark & \xmark & \xmark & \xmark & \xmark & \xmark & \xmark & \cmark & \cmark & \xmark & \xmark & \xmark \\
					CONFINE~\cite{CONFINE} & RAID'20 & C2P & \xmark & \xmark & \xmark & \xmark & \cmark & \cmark & \xmark & \xmark & \cmark & \xmark & \xmark & \cmark & \cmark & \xmark & \xmark & \xmark \\
					
					CARVE~\cite{CARVE} & FEAST'19 & S2S & \xmark & \xmark & \cmark & \cmark & \cmark & \xmark & \xmark & \cmark & \xmark & \cmark & \xmark & \cmark & \cmark & \xmark & \xmark & \xmark \\
					BinRec~\cite{BinRec} & FEAST'18 & B2B & \xmark & \cmark & \cmark & \cmark & \cmark & \cmark & \xmark & \cmark & \xmark & \cmark & \cmark & \cmark & \cmark & \xmark & \xmark & \xmark \\
					
					Nibbler~\cite{Nibbler} & ACSAC'19 & B2B & \xmark & \cmark & \xmark & \xmark & \cmark & \xmark & \xmark & \xmark & \xmark & \xmark & \cmark & \cmark & \xmark & \cmark & \xmark & \xmark \\
					
					JShrink~\cite{JShrink} & FSE'20 & B2B & \cmark & \cmark & \xmark & \cmark & \cmark & \cmark & \xmark & \xmark & \cmark & \cmark & \xmark & \cmark & \cmark & \xmark & \xmark & \xmark \\
					JReduce~\cite{JReduce} & FSE'19 & B2B & \cmark & \xmark & \xmark & \xmark & \xmark & \cmark & \xmark & \xmark & \cmark & \xmark & \xmark & \xmark & \xmark & \cmark & \xmark & \xmark \\
					Cimplifier~\cite{Cimplifier} & FSE'17 & C2C & \cmark & \xmark & \xmark & \xmark & \cmark & \cmark & \xmark & \xmark & \cmark & \xmark & \xmark & \cmark & \cmark & \cmark & \xmark & \xmark \\
					Picup~\cite{Picup} & FSE'23 & S2B & \xmark & \cmark & \xmark & \xmark & \cmark & \cmark & \cmark & \cmark & \cmark & \cmark & \xmark & \cmark & \cmark & \xmark & \xmark & \xmark \\
					
					Minimon~\cite{minimon} & ICSE'24 & B2B & \xmark & \cmark & \xmark & \xmark & \cmark & \cmark & \cmark & \xmark & \xmark & \cmark & \xmark & \xmark & \cmark & \xmark & \xmark & \xmark \\
					Perses~\cite{Perses} & ICSE'18 & S2S & \xmark & \cmark & \cmark & \cmark & \cmark & \cmark & \xmark & \xmark & \cmark & \xmark & \xmark & \xmark & \xmark & \cmark & \xmark & \xmark \\
					
					AutoDebloater~\cite{AutoDebloater} & ASE'23 & B2B & \cmark & \cmark & \xmark & \xmark & \cmark & \cmark & \xmark & \xmark & \cmark & \xmark & \xmark & \xmark & \xmark & \xmark & \xmark & \xmark \\
					DomGad~\cite{DomGad} & ASE'20 & S2S & \xmark & \xmark & \cmark & \cmark & \cmark & \cmark & \xmark & \cmark & \xmark & \cmark & \xmark & \cmark & \cmark & \cmark & \xmark & \xmark \\
					
					BlankIt~\cite{BlankIt} & PLDI'20 & B2B & \xmark & \cmark & \xmark & \xmark & \cmark & \cmark & \cmark & \xmark & \xmark & \cmark & \xmark & \cmark & \xmark & \cmark & \xmark & \xmark \\
					C-Reduce~\cite{C-Reduce} & PLDI'12 & S2S & \xmark & \cmark & \cmark & \cmark & \cmark & \xmark & \xmark & \xmark & \cmark & \xmark & \cmark & \xmark & \cmark & \xmark & \xmark & \xmark \\
					
					Decker~\cite{Decker} & ASPLOS'23 & S2B & \xmark & \cmark & \xmark & \xmark & \cmark & \cmark & \xmark & \cmark & \xmark & \cmark & \xmark & \cmark & \xmark & \cmark & \xmark & \xmark \\
					$\mu$Trimmer~\cite{mTrimmer} & ASPLOS'22 & B2B & \xmark & \cmark & \cmark & \xmark & \cmark & \xmark & \xmark & \xmark & \xmark & \xmark & \xmark & \cmark & \xmark & \xmark & \xmark & \xmark \\
					
					Trimmer~\cite{Trimmer} & TSE'22 & S2B & \xmark & \cmark & \cmark & \cmark & \cmark & \xmark & \xmark & \cmark & \xmark & \cmark & \xmark & \cmark & \cmark & \cmark & \xmark & \xmark \\
					XDebloat~\cite{XDebloat} & TSE'21 & B2B & \cmark & \cmark & \cmark & \cmark & \cmark & \xmark & \xmark & \cmark & \xmark & \cmark & \cmark & \xmark & \cmark & \cmark & \xmark & \xmark \\
					NA~\cite{redLoadTime} & TSE'21 & S2S & \xmark & \cmark & \cmark & \cmark & \xmark & \cmark & \xmark & \xmark & \cmark & \xmark & \xmark & \xmark & \xmark & \xmark & \xmark & \xmark \\
					
					BLADE~\cite{BLADE} & SecDev'23 & S2S & \xmark & \cmark & \cmark & \cmark & \cmark & \cmark & \xmark & \xmark & \cmark & \xmark & \xmark & \cmark & \cmark & \cmark & \xmark & \xmark \\
					JDBL~\cite{JDBL} & \shortstack{Trans. SE. Meth.'23} & S2B & \cmark & \cmark & \xmark & \xmark & \xmark & \cmark & \xmark & \xmark & \cmark & \cmark & \xmark & \xmark & \cmark & \xmark & \xmark & \xmark \\
					OCCAM~\cite{OCCAM} & Commun. ACM'23 & S2B & \xmark & \cmark & \cmark & \cmark & \cmark & \xmark & \xmark & \cmark & \xmark & \xmark & \xmark & \cmark & \cmark & \cmark & \xmark & \xmark \\
					
					Ancile~\cite{Ancile} & CODASPY '21 & S2B & \xmark & \cmark & \xmark & \xmark & \cmark & \cmark & \xmark & \cmark & \xmark & \cmark & \xmark & \cmark & \cmark & \cmark & \xmark & \xmark \\
					JSLIM~\cite{JSLIM} & EISA 2021 & S2S & \xmark & \cmark & \xmark & \xmark & \cmark & \xmark & \cmark & \xmark & \xmark & \xmark & \xmark & \cmark & \cmark & \xmark & \xmark & \xmark \\
					PRAT~\cite{PRAT} & TOSEM'21 & S2B & \xmark & \cmark & \xmark & \cmark & \cmark & \cmark & \xmark & \cmark & \cmark & \xmark & \xmark & \cmark & \cmark & \xmark & \xmark & \xmark \\
					
					DEPCLEAN~\cite{DEPCLEAN} & \shortstack{Empir SE'21} & D2D & \cmark & \xmark & \xmark & \xmark & \cmark & \xmark & \xmark & \cmark & \xmark & \xmark & \xmark & \xmark & \xmark & \xmark & \xmark & \xmark \\
					
					DECAF~\cite{DECAF} & ICSE-SEIP'20 & B2B & \cmark & \xmark & \xmark & \xmark & \cmark & \cmark & \xmark & \cmark & \cmark & \cmark & \cmark & \cmark & \cmark & \xmark & \xmark & \xmark \\
					
					NA~\cite{Configuration-Driven} & EuroSec'19 & S2B & \cmark & \cmark & \xmark & \xmark & \cmark & \cmark & \xmark & \cmark & \xmark & \cmark & \xmark & \cmark & \xmark & \xmark & \xmark & \xmark \\
					DeepOCCAM~\cite{DeepOCCAM} & MLforSystems'19 & S2B & \xmark & \cmark & \cmark & \cmark & \cmark & \xmark & \cmark & \cmark & \xmark & \xmark & \xmark & \cmark & \xmark & \cmark & \xmark & \xmark \\
					BINTRIMMER~\cite{BINTRIMMER} & LNSC'19 & B2B & \xmark & \cmark & \cmark & \xmark & \cmark & \xmark & \xmark & \xmark & \xmark & \cmark & \xmark & \cmark & \cmark & \cmark & \xmark & \xmark \\ 
					
					RedDroid~\cite{RedDroid} & ISSRE'18 & B2B & \xmark & \cmark & \xmark & \xmark & \cmark & \xmark & \xmark & \xmark & \xmark & \xmark & \xmark & \xmark & \xmark & \xmark & \xmark & \xmark \\
					
					SPEAKER~\cite{speaker} & DIMVA'17 & C2P & \xmark & \xmark & \xmark & \xmark & \xmark & \cmark & \xmark & \xmark & \cmark & \xmark & \cmark & \cmark & \xmark & \xmark & \xmark & \xmark \\
					
					Jred~\cite{jred} & COMPSAC'16 & B2B & \cmark & \cmark & \xmark & \xmark & \cmark & \xmark & \xmark & \xmark & \xmark & \xmark & \cmark & \cmark & \xmark & \cmark & \xmark & \xmark \\
					
					NA~\cite{Energy} & ISLPED '01 & S2S & \xmark & \cmark & \cmark & \cmark & \cmark & \cmark & \xmark & \cmark & \xmark & \cmark & \xmark & \xmark & \cmark & \cmark & \xmark & \cmark \\ 
					\hline
				\end{tabular}
			}
		}
		\label{table:summary}
	\end{table*}

\end{landscape}

\subsubsection{Functionality}

This component presents three strategies to identify unneeded functionalities at a high level. 
\begin{itemize}
    \item \textbf{Configuration.} In this strategy, the debloating workflow receives program configurations as input, which are to be preserved in the debloated output. These configurations may also specify particular points of interest, such as specific functions and libraries. For example, LMCAS~\cite{LMCAS} requires configurations via command-line arguments or a configuration file, mirroring the program's standard execution approach. Conversely, tools like OCCAM~\cite{OCCAM} and Trimmer~\cite{Trimmer} use a template format to input the required configurations. Other tools, like temporal-specialization~\cite{temporal}, anticipate the configuration in the form of a list of key functions from the input artifact.
    \item  \textbf{Test cases.} This debloating strategy requires a collection of test cases to represent the program's usage profile post-debloating. Tools like Chisel~\cite{CHISEL} and Razor~\cite{RAZOR} use test cases supplied by the developers as input. Other tools, such as Ancile~\cite{Ancile}, employ fuzzing techniques to generate these test cases. Hacksaw~\cite{Hacksaw} utilizes hardware probing to identify necessary device drivers to perform kernel debloating.
    \item  \textbf{Annotation.} In this strategy, the input program is augmented with specific logic. This addition is either to gather particular information during dynamic analysis, such as profiling, or to initiate different actions. For instance, LMCAS~\cite{LMCAS} marks specific locations in the program to signal the completion of the profiling process. Conversely, Slimium~\cite{Slimium} employs binary instrumentation to track functions that are called during runtime. 
\end{itemize}

\begin{figure}[h]
	\begin{minipage}[b]{0.48\linewidth}
		\centering
		\includegraphics[width=\linewidth]{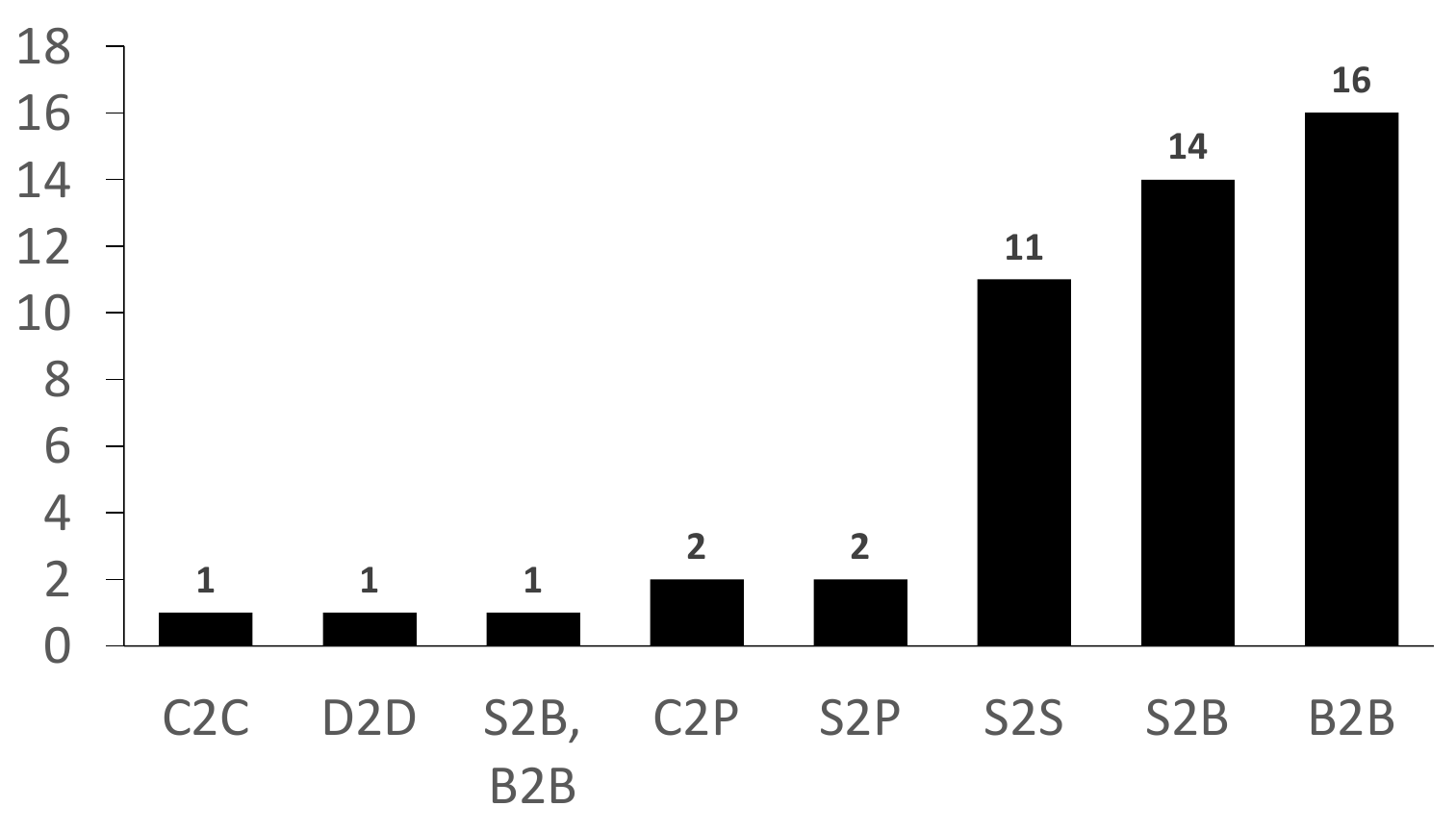}
		%\vspace{-0.3cm}
		\caption{I/O artifacts type mappings across tools.}
		%\vspace{-0.3cm}
		\label{fig:artifacts_count}
	\end{minipage}
	\hfill
	\begin{minipage}[b]{0.48\linewidth}
		\centering
		\includegraphics[width=\linewidth]{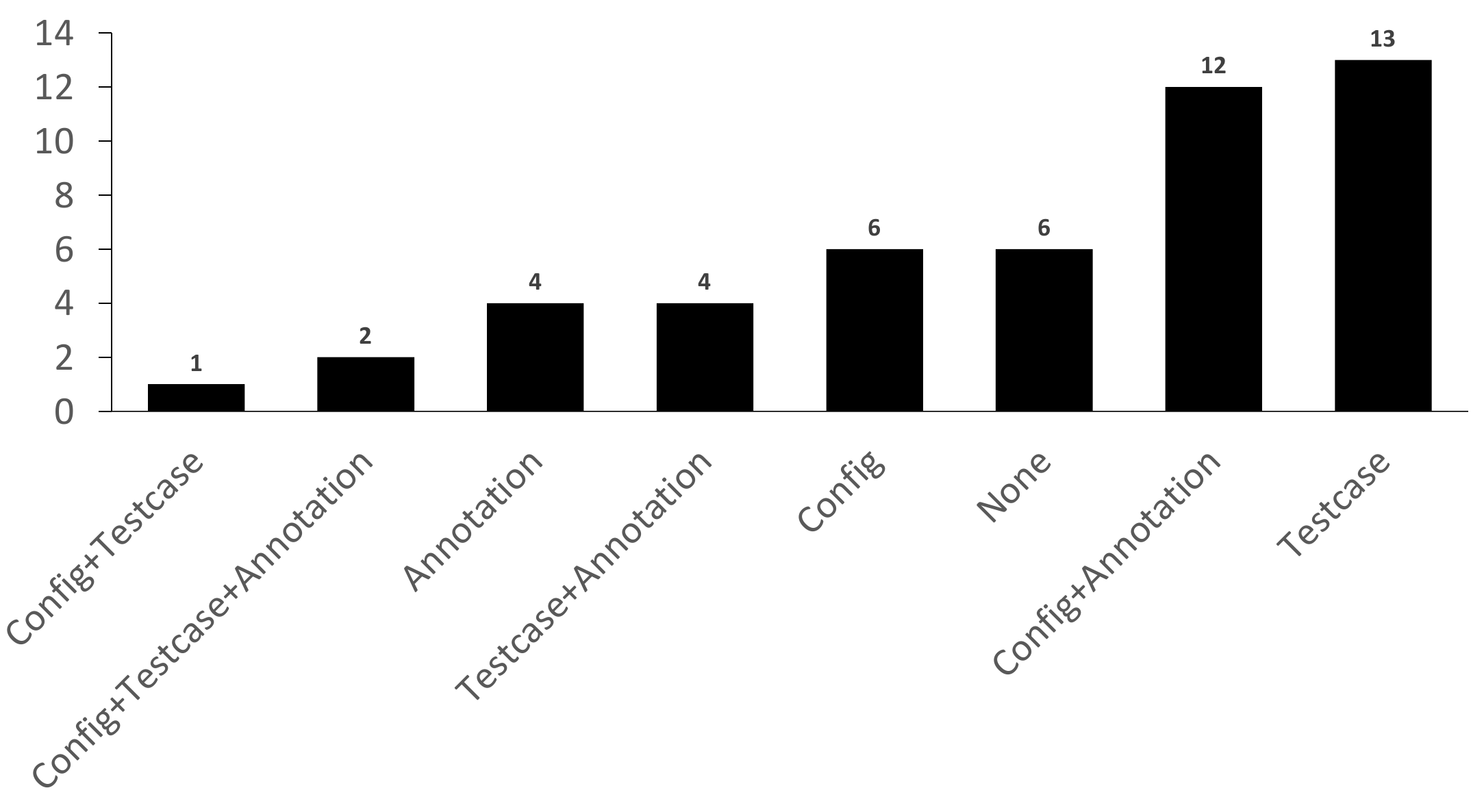}
		%\vspace{-0.3cm}
		\caption{Strategies to identify functionality across tools.}
		%\vspace{-0.3cm}
		\label{fig:functionality_count}
	\end{minipage}
\end{figure}

\noindent
Six tools~\cite{Nibbler, Mininode, RedDroid,jred, temporal, JSLIM} (under the \texttt{none} category in Figure~\ref{fig:functionality_count}) depend solely on static analysis techniques to pinpoint unneeded functionalities, eliminating the need for explicit expression of these functionalities. Thus, we consider functionality as an optional component of the debloating strategy stage in the workflow. In particular, all these tools use only static analysis and identify unused code by performing a reachability analysis on call graphs~\cite{Nibbler,RedDroid} or dependency graphs~\cite{Mininode}.

\subsubsection{Analysis}

This component describes program analysis techniques that have been utilized by various software debloating tools.  
\begin{itemize}
    \item \textbf{Static Analysis.} This analysis focuses on building various types of graphs, such as call graphs, Control Flow Graphs (CFGs), and dependency graphs, to identify dependencies at multiple levels of granularity. C2C~\cite{C2C} generates a CFG and performs data flow analysis during its analysis. 
    % Within the scope of container debloating tools, static analysis mainly involves examining traces gathered from dynamic analysis. 
    Additionally, an important aspect of static analysis is the optimization and elimination of unnecessary dependencies. For example, LMCAS~\cite{LMCAS}, OCCAM~\cite{OCCAM}, and Trimmer~\cite{Trimmer} implement LLVM passes to simplify and remove unneeded code. 
    \item \textbf{Dynamic Analysis.} In this analysis technique, run-time data is collected to identify essential dependencies that must be preserved. This technique typically involves instrumenting the application before execution. Various tools have been used to aid in dynamic analysis. For example, LMCAS~\cite{LMCAS} and LightBlue~\cite{LightBlue} employ symbolic execution, whereas other tools such as Slimium~\cite{Slimium} have developed their own dynamic analysis methods. 
    % \textbf{@Ashish: should we consider OCCAM uses dynamic analysis?}
    \item  \textbf{ML-Assisted Static/Dynamic Analysis.} Machine Learning (ML), distinct from prior techniques, is often employed in conjunction with program analysis. An example of this is Chisel~\cite{CHISEL}, which combines delta debugging with reinforcement learning.
\end{itemize}

\begin{figure}[h]
	\begin{minipage}[b]{0.48\linewidth}
		\centering
		\includegraphics[width=\linewidth]{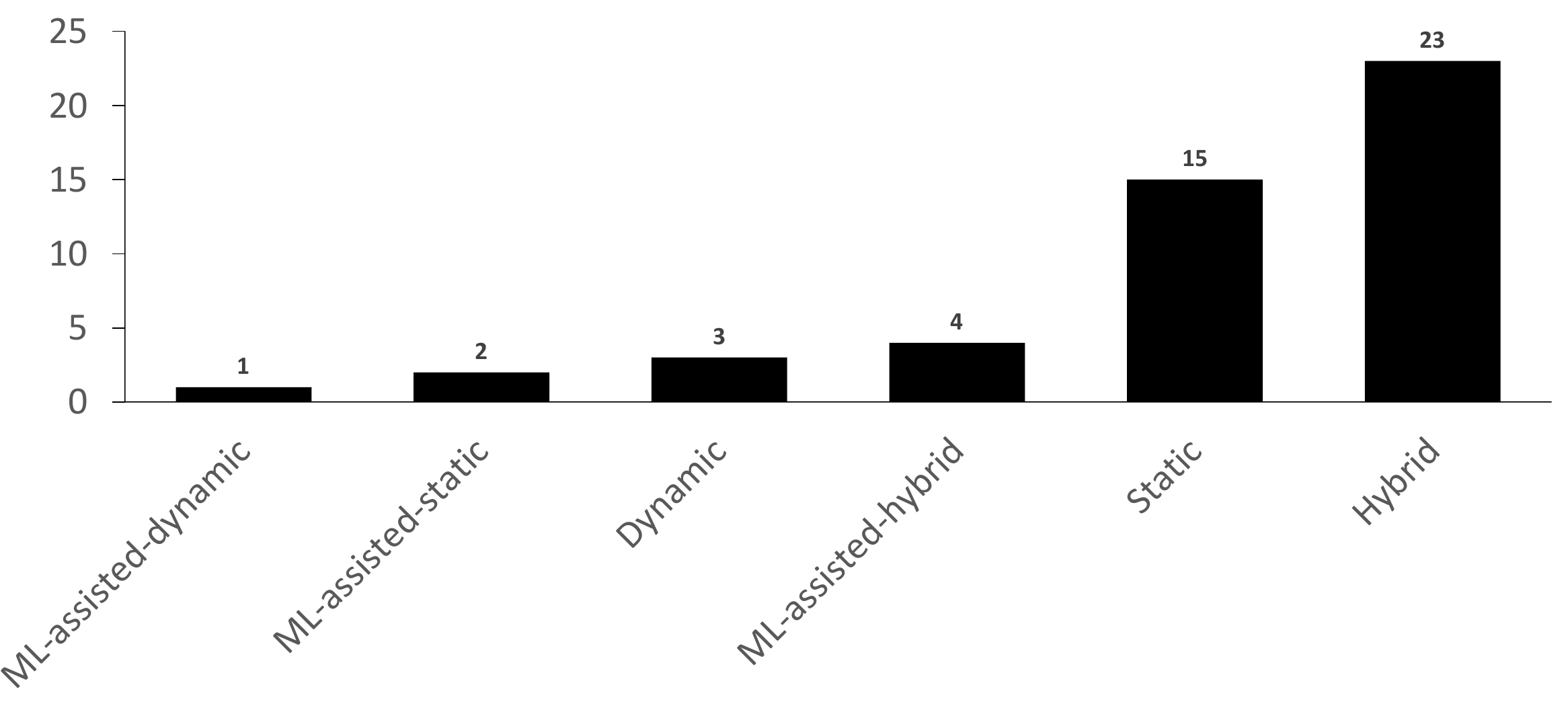}
		%\vspace{-0.3cm}
		\caption{Analysis techniques across tools.}
		\label{fig:analysis_count}
	\end{minipage}
	\hfill
	\begin{minipage}[b]{0.48\linewidth}
		\centering
		\includegraphics[width=\linewidth]{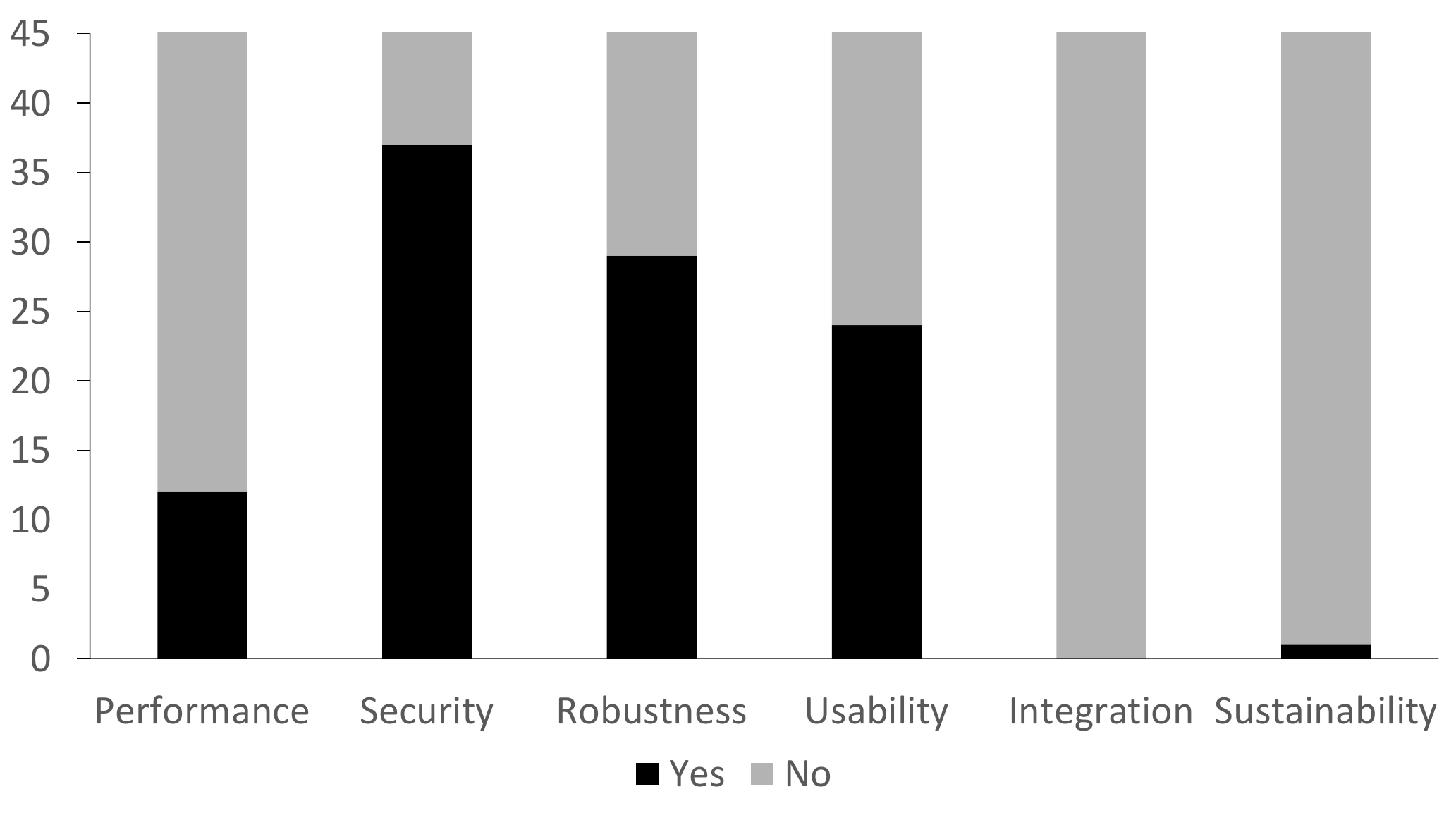}
		%\vspace{-0.3cm}
		\caption{Evaluation criteria across tools.}
		\label{fig:evaluation_count}
	\end{minipage}
\end{figure}

\noindent
A variety of tools have utilized a blend of static and dynamic analyses, sometimes supplemented with machine learning (ML). For instance, Confine~\cite{CONFINE} and Piece-Wise~\cite{Piece-Wise} employ hybrid analysis techniques for debloating containers and libraries. BlankIt~\cite{BlankIt}, another hybrid analysis tool, focuses on debloating shared libraries and incorporates ML, specifically decision trees, to predict the functions required at a particular call site during execution. Figure~\ref{fig:analysis_count} presents the number of debloating tools that fall under the different analysis categories.

\subsubsection{Removal Granularity}
Software debloating tools aim to eliminate unnecessary code and dependencies, but they do so at different levels of granularity. %This section explores the various levels of removal employed by debloating tools. 
As shown in Figure~\ref{fig:taxonomy}, there are four distinct levels of removal granularity in the context of software debloating: (1) instruction or statement, (2) basic block, (3) function or library, and (4) file, including class or dependency management. Notably, some tools, such as Confine~\cite{CONFINE}, temporal-specialization~\cite{temporal}, and SPEAKER~\cite{speaker}, primarily aim to reduce syscalls rather than directly removing code elements.

%\vspace{-0.3cm}
\subsection{Evaluation Criteria}

In this stage of the workflow, measurable metrics are applied to the artifact before and after debloating to assess its effectiveness from multiple perspectives. As shown in Figure~\ref{fig:evaluation_count}, the following are the main evaluation criteria used in the reviewed literature:
\begin{itemize}
    \item \textbf{Performance.}
This metric evaluates the performance of the debloated program in terms of its memory usage, CPU utilization, bandwidth, and runtime.

    \item \textbf{Security.}
Tools for software debloating, particularly those created by the security community, are designed primarily to improve security and minimize potential attack vectors. Their security assessment predominantly revolves around quantifying the count of Common Vulnerability Exposures (CVEs) and gadgets.

    \item \textbf{Robustness.}
This metric is analyzed from various viewpoints: \textit{correctness} and \textit{generality}. The latter evaluates how accurately a debloated program functions with inputs that were not part of the original usage profile~\cite{Generality}. Methods like fuzzing and test cases are used to assess the correctness. Tools like LMCAS~\cite{LMCAS} and Razor~\cite{RAZOR} also examine for undesirable behaviors, including incorrect operations, infinite loops, crashes, and missing output.

    \item \textbf{Usability.}
This metric focuses on assessing the resources needed by the debloating tool (not the debloated artifact), examined from the perspective of runtime and functionality requirements. For example, Chisel~\cite{CHISEL} utilizes reinforcement learning along with delta debugging, thus increasing the overhead of running it.
% necessitating the creation of test cases that reflect the essential functionality. This requirement contributes to Chisel's long running time.

\item \textbf{Integration.}
BLADE~\cite{BLADE} views software debloating as essential for ecosystems such as clouds, requiring rapid analysis to support integration with continuous integration and continuous delivery (CI/CD) infrastructures. Consequently, this metric evaluates the capacity of debloating tools to integrate with established ecosystem infrastructures. Despite BLADE's vision, its evaluation did not encompass demonstrating integration capabilities.  

\item \textbf{Sustainability.}
This metric evaluates the quality of debloated programs based on carbon footprint and energy reduction. We found only one debloating tool~\cite{Energy} that primarily targets energy reduction and thus focuses on only evaluating this factor.
\end{itemize}
\vspace{-0.3cm}
%\section{Open Problems and Future Research}
\section{Future Research} \label{sec:challenges}
This section presents open problems in software debloating and calls for solutions that are practical, usable, and secure.

\vspace{-0.2cm}
\subsection{Software Robustness}
Software debloating tools typically prioritize the preservation of error-free paths by utilizing test cases that reflect the intended behavior or providing accurate configurations. As a result, event handler procedures can be removed from the debloated programs, affecting the reliability and robustness of the application. Ancile~\cite{Ancile} includes the reachable exception handlers in
the final binary.
Carve~\cite{CARVE} avoids introducing vulnerabilities by replacing debloated code with replacement code that preserves high-level program properties. In some cases, during the debloating process, Carve replaces the \textit{switch} block with exception handling code that traps execution before code blocks that become vulnerable after debloating. However, more work is needed to balance robustness and removal~\cite{blafs}.

\vspace{-0.2cm}
\subsection{SBOM Generation}
%Software debloating is key to generating accurate Software Bills of Materials (SBOMs) 
The generation of Software Bills of Materials (SBOMs) has gained significant importance as regulatory bodies like the US National Telecommunications and Information Administration (NTIA) mandate the disclosure of primary and transitive dependencies, thereby documenting the entire code provenance~\cite{ntia}. 
MMLB~\cite{mmlb} constructs dependency trees for ML containers to investigate the impact of debloating on the number of direct and transitive dependencies.
Recent dependency management approaches, such as DepsRAG, advocate the use of large language models (LLMs) and knowledge graphs (KGs) to support the generation of SBOMs~\cite{alhanahnah2024depesrag}. Identifying software dependencies constitutes a fundamental aspect of the debloating process,
% to eliminate unnecessary dependencies, 
positioning it as a potential facilitator for SBOM generation. The intersection highlights the necessity for further research in this domain.

\vspace{-0.2cm}
\subsection{ML for Debloating}
Our investigation indicates that only a limited number of tools (7 out of 48) utilize machine learning (ML) to support debloating. Given the widespread adoption of ML, particularly LLMs, in tasks such as code generation and program repair, there is a compelling need to explore how LLMs can enhance the debloating process.   

\vspace{-0.2cm}
\subsection{Debloating Impact on Sustainability}
In our literature review, we found only one debloating tool~\cite{Energy} specifically designed to reduce energy consumption. This underscores the need for increased focus and effort in this area. Consequently, we consider this to be an open problem that is worth investigating, especially if new debloating methods can significantly decrease energy use and, as a result, cut down on carbon emissions. Subsequently, researchers might investigate the creation of debloating-driven methods aimed at eliminating software dependencies to achieve energy savings.

\vspace{-0.2cm}
\subsection{CI/CD Integration}
Software debloating has often been approached in a siloed manner, which has limited its widespread adoption in real-world scenarios. In today's Software Development Lifecycle (SDLC) and software supply chains, there is a focus on transparency and automation, incorporating practices like CI/CD. CI involves regularly merging code changes from various developers into a central repository, often multiple times per day. CD ensures that the code in the repository is always ready for release, having passed automated tests and quality assessments. Consequently, there are several challenges to address for integrating software debloating tools into CI/CD pipelines~\cite{BLADE}. For example, determining which test cases should validate a release that includes a debloated version of the application and deciding the necessary security analyses are key considerations. 
% \ma{@Ashish: do you have any reference to support/motivate this?}
\vspace{-0.44cm}
\section{Conclusion}
Software debloating is an essential dependency management approach for enhancing both security and performance by removing unnecessary code from applications. Our SoK highlights the diverse techniques and tools available, identifies significant advancements, and points out continuing challenges. We provide a foundational reference, aiming to guide future research and improvements in software debloating.
\vspace{-0.2cm}

\section*{Acknowledgement}

This material is based upon work supported by the National Science Foundation (NSF) under Grant ACI-1440800 and the Office of Naval Research (ONR) under Contracts N68335-17-C-0558 and N00014-24-1-2049. Any opinions, findings, conclusions, or recommendations expressed in this material are those of the authors and do not necessarily reflect the views of NSF or ONR.

%\scriptsize
\bibliographystyle{ACM-Reference-Format}
\bibliography{ref,Gehani}


\begin{thebibliography}{63}

%%% the Web address.

\ifx \showCODEN    \undefined \def \showCODEN     #1{\unskip}     \fi
\ifx \showDOI      \undefined \def \showDOI       #1{#1}\fi
\ifx \showISBNx    \undefined \def \showISBNx     #1{\unskip}     \fi
\ifx \showISBNxiii \undefined \def \showISBNxiii  #1{\unskip}     \fi
\ifx \showISSN     \undefined \def \showISSN      #1{\unskip}     \fi
\ifx \showLCCN     \undefined \def \showLCCN      #1{\unskip}     \fi
\ifx \shownote     \undefined \def \shownote      #1{#1}          \fi
\ifx \showarticletitle \undefined \def \showarticletitle #1{#1}   \fi
\ifx \showURL      \undefined \def \showURL       {\relax}        \fi
% invisible to TeX
\providecommand\bibfield[2]{#2}
\providecommand\bibinfo[2]{#2}
\providecommand\natexlab[1]{#1}
\providecommand\showeprint[2][]{arXiv:#2}

\bibitem[TPC({[n.\,d.]})]%
        {TPCP}
 \bibinfo{year}{[n.\,d.]}\natexlab{}.
\newblock \bibinfo{title}{defense.gov}.
\newblock
  \bibinfo{howpublished}{\url{https://media.defense.gov/2020/May/18/2002302043/-1/-1/1/NPG17.PDF}}.
\newblock
\newblock
\shownote{[Accessed 15-06-2024]}.


\bibitem[nti({[n.\,d.]})]%
        {ntia}
 \bibinfo{year}{[n.\,d.]}\natexlab{}.
\newblock \bibinfo{title}{ntia.doc.gov}.
\newblock
  \bibinfo{howpublished}{\url{https://www.ntia.doc.gov/files/ntia/publications/sbom_minimum_elements_report.pdf}}.
\newblock
\newblock
\shownote{[Accessed 15-06-2024]}.


\bibitem[Agadakos et~al\mbox{.}(2019)]%
        {Nibbler}
\bibfield{author}{\bibinfo{person}{Ioannis Agadakos}, \bibinfo{person}{Di Jin},
  \bibinfo{person}{David Williams-King}, \bibinfo{person}{Vasileios~P.
  Kemerlis}, {and} \bibinfo{person}{Georgios Portokalidis}.}
  \bibinfo{year}{2019}\natexlab{}.
\newblock \showarticletitle{Nibbler: Debloating Binary Shared Libraries}. In
  \bibinfo{booktitle}{\emph{35th Annual Computer Security Applications
  Conference}} (San Juan, Puerto Rico, USA) \emph{(\bibinfo{series}{ACSAC
  '19})}. \bibinfo{publisher}{Association for Computing Machinery},
  \bibinfo{address}{New York, NY, USA}, \bibinfo{pages}{70–83}.
\newblock
\showISBNx{9781450376280}
\urldef\tempurl%
\url{https://doi.org/10.1145/3359789.3359823}
\showDOI{\tempurl}


\bibitem[Ahmad et~al\mbox{.}(2022)]%
        {Trimmer}
\bibfield{author}{\bibinfo{person}{Aatira~Anum Ahmad},
  \bibinfo{person}{Abdul~Rafae Noor}, \bibinfo{person}{Hashim Sharif},
  \bibinfo{person}{Usama Hameed}, \bibinfo{person}{Shoaib Asif},
  \bibinfo{person}{Mubashir Anwar}, \bibinfo{person}{Ashish Gehani},
  \bibinfo{person}{Fareed Zaffar}, {and} \bibinfo{person}{Junaid~Haroon
  Siddiqui}.} \bibinfo{year}{2022}\natexlab{}.
\newblock \showarticletitle{Trimmer: An Automated System for
  Configuration-Based Software Debloating}.
\newblock \bibinfo{journal}{\emph{IEEE Transactions on Software Engineering}}
  \bibinfo{volume}{48}, \bibinfo{number}{9} (\bibinfo{year}{2022}),
  \bibinfo{pages}{3485--3505}.
\newblock
\urldef\tempurl%
\url{https://doi.org/10.1109/TSE.2021.3095716}
\showDOI{\tempurl}


\bibitem[Alhanahnah et~al\mbox{.}(2024)]%
        {alhanahnah2024depesrag}
\bibfield{author}{\bibinfo{person}{Mohannad Alhanahnah}, \bibinfo{person}{Yazan
  Boshmaf}, {and} \bibinfo{person}{Benoit Baudry}.}
  \bibinfo{year}{2024}\natexlab{}.
\newblock \showarticletitle{DepesRAG: Towards Managing Software Dependencies
  using Large Language Models}.
\newblock \bibinfo{journal}{\emph{arXiv preprint arXiv:2405.20455}}
  (\bibinfo{year}{2024}).
\newblock


\bibitem[Alhanahnah et~al\mbox{.}(2022)]%
        {LMCAS}
\bibfield{author}{\bibinfo{person}{Mohannad Alhanahnah},
  \bibinfo{person}{Rithik Jain}, \bibinfo{person}{Vaibhav Rastogi},
  \bibinfo{person}{Somesh Jha}, {and} \bibinfo{person}{Thomas Reps}.}
  \bibinfo{year}{2022}\natexlab{}.
\newblock \showarticletitle{Lightweight, Multi-Stage, Compiler-Assisted
  Application Specialization}. In \bibinfo{booktitle}{\emph{2022 IEEE 7th
  European Symposium on Security and Privacy (EuroS\&P)}}.
  \bibinfo{pages}{251--269}.
\newblock
\urldef\tempurl%
\url{https://doi.org/10.1109/EuroSP53844.2022.00024}
\showDOI{\tempurl}


\bibitem[Ali et~al\mbox{.}(2023a)]%
        {BLADE}
\bibfield{author}{\bibinfo{person}{Muaz Ali}, \bibinfo{person}{Rumaisa Habib},
  \bibinfo{person}{Ashish Gehani}, \bibinfo{person}{Sazzadur Rahaman}, {and}
  \bibinfo{person}{Zartash Uzmi}.} \bibinfo{year}{2023}\natexlab{a}.
\newblock \showarticletitle{Blade: Scalable Source Code Debloating Framework}.
  In \bibinfo{booktitle}{\emph{2023 IEEE Secure Development Conference
  (SecDev)}}.
\newblock


\bibitem[Ali et~al\mbox{.}(2023b)]%
        {sok_ESORICS}
\bibfield{author}{\bibinfo{person}{Muaz Ali}, \bibinfo{person}{Muhammad
  Muzammil}, \bibinfo{person}{Faraz Karim}, \bibinfo{person}{Ayesha Naeem},
  \bibinfo{person}{Rukhshan Haroon}, \bibinfo{person}{Muhammad Haris},
  \bibinfo{person}{Huzaifah Nadeem}, \bibinfo{person}{Waseem Sabir},
  \bibinfo{person}{Fahad Shaon}, \bibinfo{person}{Fareed Zaffar},
  {et~al\mbox{.}}} \bibinfo{year}{2023}\natexlab{b}.
\newblock \showarticletitle{SoK: A Tale of Reduction, Security, and
  Correctness-Evaluating Program Debloating Paradigms and Their Compositions}.
  ESORICS.
\newblock


\bibitem[Biswas et~al\mbox{.}(2021)]%
        {Ancile}
\bibfield{author}{\bibinfo{person}{Priyam Biswas}, \bibinfo{person}{Nathan
  Burow}, {and} \bibinfo{person}{Mathias Payer}.}
  \bibinfo{year}{2021}\natexlab{}.
\newblock \showarticletitle{Code Specialization through Dynamic Feature
  Observation}. In \bibinfo{booktitle}{\emph{11th ACM Conference on Data and
  Application Security and Privacy}} (Virtual Event, USA)
  \emph{(\bibinfo{series}{CODASPY '21})}. \bibinfo{publisher}{Association for
  Computing Machinery}, \bibinfo{address}{New York, NY, USA},
  \bibinfo{pages}{257–268}.
\newblock
\showISBNx{9781450381437}
\urldef\tempurl%
\url{https://doi.org/10.1145/3422337.3447844}
\showDOI{\tempurl}


\bibitem[Brown et~al\mbox{.}(2024)]%
        {sok_usenix}
\bibfield{author}{\bibinfo{person}{Michael~D. Brown}, \bibinfo{person}{Adam
  Meily}, \bibinfo{person}{Brian Fairservice}, \bibinfo{person}{Akshay Sood},
  \bibinfo{person}{Jonathan Dorn}, \bibinfo{person}{Eric Kilmer}, {and}
  \bibinfo{person}{Ronald Eytchison}.} \bibinfo{year}{2024}\natexlab{}.
\newblock \bibinfo{title}{A Broad Comparative Evaluation of Software Debloating
  Tools}.
\newblock
\newblock
\showeprint[arxiv]{2312.13274}


\bibitem[Brown and Pande(2019)]%
        {CARVE}
\bibfield{author}{\bibinfo{person}{Michael~D. Brown} {and}
  \bibinfo{person}{Santosh Pande}.} \bibinfo{year}{2019}\natexlab{}.
\newblock \showarticletitle{CARVE: Practical Security-Focused Software
  Debloating Using Simple Feature Set Mappings}. In
  \bibinfo{booktitle}{\emph{3rd ACM Workshop on Forming an Ecosystem Around
  Software Transformation}} (London, United Kingdom)
  \emph{(\bibinfo{series}{FEAST'19})}. \bibinfo{publisher}{Association for
  Computing Machinery}, \bibinfo{address}{New York, NY, USA},
  \bibinfo{pages}{1–7}.
\newblock
\showISBNx{9781450368346}
\urldef\tempurl%
\url{https://doi.org/10.1145/3338502.3359764}
\showDOI{\tempurl}


\bibitem[Bruce et~al\mbox{.}(2020)]%
        {JShrink}
\bibfield{author}{\bibinfo{person}{Bobby~R. Bruce}, \bibinfo{person}{Tianyi
  Zhang}, \bibinfo{person}{Jaspreet Arora}, \bibinfo{person}{Guoqing~Harry Xu},
  {and} \bibinfo{person}{Miryung Kim}.} \bibinfo{year}{2020}\natexlab{}.
\newblock \showarticletitle{JShrink: In-Depth Investigation into Debloating
  Modern Java Applications}. In \bibinfo{booktitle}{\emph{28th ACM Joint
  Meeting on European Software Engineering Conference and Symposium on the
  Foundations of Software Engineering}} (Virtual Event, USA)
  \emph{(\bibinfo{series}{ESEC/FSE 2020})}. \bibinfo{publisher}{Association for
  Computing Machinery}, \bibinfo{address}{New York, NY, USA},
  \bibinfo{pages}{135–146}.
\newblock
\showISBNx{9781450370431}
\urldef\tempurl%
\url{https://doi.org/10.1145/3368089.3409738}
\showDOI{\tempurl}


\bibitem[Christensen et~al\mbox{.}(2020)]%
        {DECAF}
\bibfield{author}{\bibinfo{person}{Jake Christensen},
  \bibinfo{person}{Ionut~Mugurel Anghel}, \bibinfo{person}{Rob Taglang},
  \bibinfo{person}{Mihai Chiroiu}, {and} \bibinfo{person}{Radu Sion}.}
  \bibinfo{year}{2020}\natexlab{}.
\newblock \showarticletitle{{DECAF}: Automatic, Adaptive De-bloating and
  Hardening of {COTS} Firmware}. In \bibinfo{booktitle}{\emph{29th USENIX
  Security Symposium (USENIX Security 20)}}. \bibinfo{publisher}{USENIX
  Association}, \bibinfo{pages}{1713--1730}.
\newblock
\showISBNx{978-1-939133-17-5}


\bibitem[Chung et~al\mbox{.}(2001)]%
        {Energy}
\bibfield{author}{\bibinfo{person}{Eui-Young Chung}, \bibinfo{person}{Luca
  Benini}, {and} \bibinfo{person}{Giovanni De~Micheli}.}
  \bibinfo{year}{2001}\natexlab{}.
\newblock \showarticletitle{Automatic Source Code Specialization for Energy
  Reduction}. In \bibinfo{booktitle}{\emph{International Symposium on Low Power
  Electronics and Design}} (Huntington Beach, California, USA)
  \emph{(\bibinfo{series}{ISLPED '01})}. \bibinfo{publisher}{Association for
  Computing Machinery}, \bibinfo{address}{New York, NY, USA},
  \bibinfo{pages}{80–83}.
\newblock
\showISBNx{1581133715}
\urldef\tempurl%
\url{https://doi.org/10.1145/383082.383099}
\showDOI{\tempurl}


\bibitem[D\"{u}sing and Hermann(2022)]%
        {vlunAnalysis}
\bibfield{author}{\bibinfo{person}{Johannes D\"{u}sing} {and}
  \bibinfo{person}{Ben Hermann}.} \bibinfo{year}{2022}\natexlab{}.
\newblock \showarticletitle{Analyzing the Direct and Transitive Impact of
  Vulnerabilities onto Different Artifact Repositories}.
\newblock \bibinfo{journal}{\emph{Digital Threats}} \bibinfo{volume}{3},
  \bibinfo{number}{4}, Article \bibinfo{articleno}{38} (\bibinfo{date}{feb}
  \bibinfo{year}{2022}), \bibinfo{numpages}{25}~pages.
\newblock
\urldef\tempurl%
\url{https://doi.org/10.1145/3472811}
\showDOI{\tempurl}


\bibitem[Farzat et~al\mbox{.}(2021)]%
        {redLoadTime}
\bibfield{author}{\bibinfo{person}{Fábio de~A. Farzat},
  \bibinfo{person}{Márcio de~O. Barros}, {and} \bibinfo{person}{Guilherme~H.
  Travassos}.} \bibinfo{year}{2021}\natexlab{}.
\newblock \showarticletitle{Evolving JavaScript Code to Reduce Load Time}.
\newblock \bibinfo{journal}{\emph{IEEE Transactions on Software Engineering}}
  \bibinfo{volume}{47}, \bibinfo{number}{8} (\bibinfo{year}{2021}),
  \bibinfo{pages}{1544--1558}.
\newblock
\urldef\tempurl%
\url{https://doi.org/10.1109/TSE.2019.2928293}
\showDOI{\tempurl}


\bibitem[Ghaffarinia and Hamlen(2019)]%
        {cf_trimming}
\bibfield{author}{\bibinfo{person}{Masoud Ghaffarinia} {and}
  \bibinfo{person}{Kevin~W. Hamlen}.} \bibinfo{year}{2019}\natexlab{}.
\newblock \showarticletitle{Binary Control-Flow Trimming}. In
  \bibinfo{booktitle}{\emph{ACM SIGSAC Conference on Computer and
  Communications Security}} (London, United Kingdom)
  \emph{(\bibinfo{series}{CCS '19})}. \bibinfo{publisher}{Association for
  Computing Machinery}, \bibinfo{address}{New York, NY, USA},
  \bibinfo{pages}{1009–1022}.
\newblock
\showISBNx{9781450367479}
\urldef\tempurl%
\url{https://doi.org/10.1145/3319535.3345665}
\showDOI{\tempurl}


\bibitem[Ghavamnia et~al\mbox{.}(2020a)]%
        {CONFINE}
\bibfield{author}{\bibinfo{person}{Seyedhamed Ghavamnia},
  \bibinfo{person}{Tapti Palit}, \bibinfo{person}{Azzedine Benameur}, {and}
  \bibinfo{person}{Michalis Polychronakis}.} \bibinfo{year}{2020}\natexlab{a}.
\newblock \showarticletitle{Confine: Automated System Call Policy Generation
  for Container Attack Surface Reduction}. In \bibinfo{booktitle}{\emph{23rd
  International Symposium on Research in Attacks, Intrusions and Defenses (RAID
  2020)}}. \bibinfo{publisher}{USENIX Association}, \bibinfo{address}{San
  Sebastian}, \bibinfo{pages}{443--458}.
\newblock
\showISBNx{978-1-939133-18-2}


\bibitem[Ghavamnia et~al\mbox{.}(2020b)]%
        {temporal}
\bibfield{author}{\bibinfo{person}{Seyedhamed Ghavamnia},
  \bibinfo{person}{Tapti Palit}, \bibinfo{person}{Shachee Mishra}, {and}
  \bibinfo{person}{Michalis Polychronakis}.} \bibinfo{year}{2020}\natexlab{b}.
\newblock \showarticletitle{Temporal System Call Specialization for Attack
  Surface Reduction}. In \bibinfo{booktitle}{\emph{29th USENIX Conference on
  Security Symposium}} \emph{(\bibinfo{series}{SEC'20})}.
  \bibinfo{publisher}{USENIX Association}, \bibinfo{address}{USA}, Article
  \bibinfo{articleno}{99}, \bibinfo{numpages}{18}~pages.
\newblock
\showISBNx{978-1-939133-17-5}


\bibitem[Ghavamnia et~al\mbox{.}(2022)]%
        {C2C}
\bibfield{author}{\bibinfo{person}{Seyedhamed Ghavamnia},
  \bibinfo{person}{Tapti Palit}, {and} \bibinfo{person}{Michalis
  Polychronakis}.} \bibinfo{year}{2022}\natexlab{}.
\newblock \showarticletitle{C2C: Fine-Grained Configuration-Driven System Call
  Filtering}. In \bibinfo{booktitle}{\emph{ACM SIGSAC Conference on Computer
  and Communications Security}} (Los Angeles, CA, USA)
  \emph{(\bibinfo{series}{CCS '22})}. \bibinfo{publisher}{Association for
  Computing Machinery}, \bibinfo{address}{New York, NY, USA},
  \bibinfo{pages}{1243–1257}.
\newblock
\showISBNx{9781450394505}
\urldef\tempurl%
\url{https://doi.org/10.1145/3548606.3559366}
\showDOI{\tempurl}


\bibitem[Hassan et~al\mbox{.}(2023)]%
        {Hassan23}
\bibfield{author}{\bibinfo{person}{Muhammad Hassan}, \bibinfo{person}{Talha
  Tahir}, \bibinfo{person}{Muhammad Farrukh}, \bibinfo{person}{Abdullah
  Naveed}, \bibinfo{person}{Anas Naeem}, \bibinfo{person}{Fahad Shaon},
  \bibinfo{person}{Fareed Zaffar}, \bibinfo{person}{Ashish Gehani}, {and}
  \bibinfo{person}{Sazzadur Rahaman}.} \bibinfo{year}{2023}\natexlab{}.
\newblock \showarticletitle{{Evaluating Container Debloaters}}.
\newblock \bibinfo{journal}{\emph{8th IEEE Secure Development Conference
  (SecDev)}} (\bibinfo{year}{2023}).
\newblock
\urldef\tempurl%
\url{https://doi.org/10.1109/SecDev56634.2023.00023}
\showURL{%
\tempurl}


\bibitem[Heo et~al\mbox{.}(2018)]%
        {CHISEL}
\bibfield{author}{\bibinfo{person}{Kihong Heo}, \bibinfo{person}{Woosuk Lee},
  \bibinfo{person}{Pardis Pashakhanloo}, {and} \bibinfo{person}{Mayur Naik}.}
  \bibinfo{year}{2018}\natexlab{}.
\newblock \showarticletitle{Effective Program Debloating via Reinforcement
  Learning}. In \bibinfo{booktitle}{\emph{ACM SIGSAC Conference on Computer and
  Communications Security}} (Toronto, Canada) \emph{(\bibinfo{series}{CCS
  '18})}. \bibinfo{publisher}{Association for Computing Machinery},
  \bibinfo{address}{New York, NY, USA}, \bibinfo{pages}{380–394}.
\newblock
\showISBNx{9781450356930}
\urldef\tempurl%
\url{https://doi.org/10.1145/3243734.3243838}
\showDOI{\tempurl}


\bibitem[Hu and Dolan-Gavitt(2022)]%
        {IRQDebloat}
\bibfield{author}{\bibinfo{person}{Zhenghao Hu} {and} \bibinfo{person}{Brendan
  Dolan-Gavitt}.} \bibinfo{year}{2022}\natexlab{}.
\newblock \showarticletitle{IRQDebloat: Reducing Driver Attack Surface in
  Embedded Devices}. In \bibinfo{booktitle}{\emph{2022 IEEE Symposium on
  Security and Privacy (SP)}}. \bibinfo{pages}{1608--1622}.
\newblock
\urldef\tempurl%
\url{https://doi.org/10.1109/SP46214.2022.9833695}
\showDOI{\tempurl}


\bibitem[Hu et~al\mbox{.}(2023)]%
        {Hacksaw}
\bibfield{author}{\bibinfo{person}{Zhenghao Hu}, \bibinfo{person}{Sangho Lee},
  {and} \bibinfo{person}{Marcus Peinado}.} \bibinfo{year}{2023}\natexlab{}.
\newblock \showarticletitle{Hacksaw: Hardware-Centric Kernel Debloating via
  Device Inventory and Dependency Analysis}. In \bibinfo{booktitle}{\emph{ACM
  SIGSAC Conference on Computer and Communications Security}}
  \emph{(\bibinfo{series}{CCS '23})}. \bibinfo{publisher}{Association for
  Computing Machinery}, \bibinfo{address}{New York, NY, USA},
  \bibinfo{pages}{1994–2008}.
\newblock
\showISBNx{9798400700507}
\urldef\tempurl%
\url{https://doi.org/10.1145/3576915.3623208}
\showDOI{\tempurl}


\bibitem[Jiang et~al\mbox{.}(2018)]%
        {RedDroid}
\bibfield{author}{\bibinfo{person}{Yufei Jiang}, \bibinfo{person}{Qinkun Bao},
  \bibinfo{person}{Shuai Wang}, \bibinfo{person}{Xiao Liu}, {and}
  \bibinfo{person}{Dinghao Wu}.} \bibinfo{year}{2018}\natexlab{}.
\newblock \showarticletitle{RedDroid: Android Application Redundancy
  Customization Based on Static Analysis}. In \bibinfo{booktitle}{\emph{2018
  IEEE 29th International Symposium on Software Reliability Engineering
  (ISSRE)}}. \bibinfo{pages}{189--199}.
\newblock
\urldef\tempurl%
\url{https://doi.org/10.1109/ISSRE.2018.00029}
\showDOI{\tempurl}


\bibitem[Jiang et~al\mbox{.}(2016)]%
        {jred}
\bibfield{author}{\bibinfo{person}{Yufei Jiang}, \bibinfo{person}{Dinghao Wu},
  {and} \bibinfo{person}{Peng Liu}.} \bibinfo{year}{2016}\natexlab{}.
\newblock \showarticletitle{JRed: Program Customization and Bloatware
  Mitigation Based on Static Analysis}. In \bibinfo{booktitle}{\emph{2016 IEEE
  40th Annual Computer Software and Applications Conference (COMPSAC)}},
  Vol.~\bibinfo{volume}{1}. \bibinfo{pages}{12--21}.
\newblock
\urldef\tempurl%
\url{https://doi.org/10.1109/COMPSAC.2016.146}
\showDOI{\tempurl}


\bibitem[Kalhauge and Palsberg(2019)]%
        {JReduce}
\bibfield{author}{\bibinfo{person}{Christian~Gram Kalhauge} {and}
  \bibinfo{person}{Jens Palsberg}.} \bibinfo{year}{2019}\natexlab{}.
\newblock \showarticletitle{Binary Reduction of Dependency Graphs}. In
  \bibinfo{booktitle}{\emph{27th ACM Joint Meeting on European Software
  Engineering Conference and Symposium on the Foundations of Software
  Engineering}} (Tallinn, Estonia) \emph{(\bibinfo{series}{ESEC/FSE 2019})}.
  \bibinfo{publisher}{Association for Computing Machinery},
  \bibinfo{address}{New York, NY, USA}, \bibinfo{pages}{556–566}.
\newblock
\showISBNx{9781450355728}
\urldef\tempurl%
\url{https://doi.org/10.1145/3338906.3338956}
\showDOI{\tempurl}


\bibitem[Koishybayev and Kapravelos(2020)]%
        {Mininode}
\bibfield{author}{\bibinfo{person}{Igibek Koishybayev} {and}
  \bibinfo{person}{Alexandros Kapravelos}.} \bibinfo{year}{2020}\natexlab{}.
\newblock \showarticletitle{Mininode: Reducing the Attack Surface of Node.js
  Applications}. In \bibinfo{booktitle}{\emph{23rd International Symposium on
  Research in Attacks, Intrusions and Defenses (RAID 2020)}}.
  \bibinfo{publisher}{USENIX Association}, \bibinfo{address}{San Sebastian},
  \bibinfo{pages}{121--134}.
\newblock
\showISBNx{978-1-939133-18-2}


\bibitem[Koo et~al\mbox{.}(2019)]%
        {Configuration-Driven}
\bibfield{author}{\bibinfo{person}{Hyungjoon Koo}, \bibinfo{person}{Seyedhamed
  Ghavamnia}, {and} \bibinfo{person}{Michalis Polychronakis}.}
  \bibinfo{year}{2019}\natexlab{}.
\newblock \showarticletitle{Configuration-Driven Software Debloating}. In
  \bibinfo{booktitle}{\emph{12th European Workshop on Systems Security}}
  (Dresden, Germany) \emph{(\bibinfo{series}{EuroSec '19})}.
  \bibinfo{publisher}{Association for Computing Machinery},
  \bibinfo{address}{New York, NY, USA}, Article \bibinfo{articleno}{9},
  \bibinfo{numpages}{6}~pages.
\newblock
\showISBNx{9781450362740}
\urldef\tempurl%
\url{https://doi.org/10.1145/3301417.3312501}
\showDOI{\tempurl}


\bibitem[Kroes et~al\mbox{.}(2018)]%
        {BinRec}
\bibfield{author}{\bibinfo{person}{Taddeus Kroes}, \bibinfo{person}{Anil
  Altinay}, \bibinfo{person}{Joseph Nash}, \bibinfo{person}{Yeoul Na},
  \bibinfo{person}{Stijn Volckaert}, \bibinfo{person}{Herbert Bos},
  \bibinfo{person}{Michael Franz}, {and} \bibinfo{person}{Cristiano
  Giuffrida}.} \bibinfo{year}{2018}\natexlab{}.
\newblock \showarticletitle{BinRec: Attack Surface Reduction Through Dynamic
  Binary Recovery}. In \bibinfo{booktitle}{\emph{Workshop on Forming an
  Ecosystem Around Software Transformation}} (Toronto, Canada)
  \emph{(\bibinfo{series}{FEAST '18})}. \bibinfo{publisher}{Association for
  Computing Machinery}, \bibinfo{address}{New York, NY, USA},
  \bibinfo{pages}{8–13}.
\newblock
\showISBNx{9781450359979}
\urldef\tempurl%
\url{https://doi.org/10.1145/3273045.3273050}
\showDOI{\tempurl}


\bibitem[Le et~al\mbox{.}(2019)]%
        {DeepOCCAM}
\bibfield{author}{\bibinfo{person}{Nham Le}, \bibinfo{person}{Ashish Gehani},
  \bibinfo{person}{Arie Gurfinkel}, \bibinfo{person}{Susmit Jha}, {and}
  \bibinfo{person}{Jorge Navas}.} \bibinfo{year}{2019}\natexlab{}.
\newblock \showarticletitle{{Reinforcement Learning Guided Software
  Debloating}}.
\newblock \bibinfo{journal}{\emph{2nd Workshop on Machine Learning for
  Systems}} (\bibinfo{year}{2019}).
\newblock


\bibitem[Lei et~al\mbox{.}(2017)]%
        {speaker}
\bibfield{author}{\bibinfo{person}{Lingguang Lei}, \bibinfo{person}{Jianhua
  Sun}, \bibinfo{person}{Kun Sun}, \bibinfo{person}{Chris Shenefiel},
  \bibinfo{person}{Rui Ma}, \bibinfo{person}{Yuewu Wang}, {and}
  \bibinfo{person}{Qi Li}.} \bibinfo{year}{2017}\natexlab{}.
\newblock \showarticletitle{SPEAKER: Split-phase execution of application
  containers}. In \bibinfo{booktitle}{\emph{Detection of Intrusions and
  Malware, and Vulnerability Assessment: 14th International Conference, DIMVA
  2017, Bonn, Germany, July 6-7, 2017, Proceedings 14}}. Springer,
  \bibinfo{pages}{230--251}.
\newblock


\bibitem[Ligatti et~al\mbox{.}(2005)]%
        {ligatti2005control}
\bibfield{author}{\bibinfo{person}{J Ligatti}, \bibinfo{person}{M Abadi},
  \bibinfo{person}{M Bidiu}, {and} \bibinfo{person}{U Erlingsson}.}
  \bibinfo{year}{2005}\natexlab{}.
\newblock \showarticletitle{Control Flow integrity}. In
  \bibinfo{booktitle}{\emph{12th ACM Conference on Computer and communications
  security}}.
\newblock


\bibitem[Liu et~al\mbox{.}(2023)]%
        {AutoDebloater}
\bibfield{author}{\bibinfo{person}{Jiakun Liu}, \bibinfo{person}{Xing Hu},
  \bibinfo{person}{Ferdian Thung}, \bibinfo{person}{Shahar Maoz},
  \bibinfo{person}{Eran Toch}, \bibinfo{person}{Debin Gao}, {and}
  \bibinfo{person}{David Lo}.} \bibinfo{year}{2023}\natexlab{}.
\newblock \showarticletitle{AutoDebloater: Automated Android App Debloating}.
  In \bibinfo{booktitle}{\emph{2023 38th IEEE/ACM International Conference on
  Automated Software Engineering (ASE)}}. IEEE, \bibinfo{pages}{2090--2093}.
\newblock


\bibitem[Liu et~al\mbox{.}(2024)]%
        {minimon}
\bibfield{author}{\bibinfo{person}{Jiakun Liu}, \bibinfo{person}{Zicheng
  Zhang}, \bibinfo{person}{Xing Hu}, \bibinfo{person}{Ferdian Thung},
  \bibinfo{person}{Shahar Maoz}, \bibinfo{person}{Debin Gao},
  \bibinfo{person}{Eran Toch}, \bibinfo{person}{Zhipeng Zhao}, {and}
  \bibinfo{person}{David Lo}.} \bibinfo{year}{2024}\natexlab{}.
\newblock \showarticletitle{MiniMon: Minimizing Android Applications with
  Intelligent Monitoring-Based Debloating}. In \bibinfo{booktitle}{\emph{2024
  IEEE/ACM 46th International Conference on Software Engineering (ICSE)}}. IEEE
  Computer Society, \bibinfo{pages}{990--990}.
\newblock


\bibitem[Malecha et~al\mbox{.}(2015)]%
        {Malecha15}
\bibfield{author}{\bibinfo{person}{Gregory Malecha}, \bibinfo{person}{Ashish
  Gehani}, {and} \bibinfo{person}{Natarajan Shankar}.}
  \bibinfo{year}{2015}\natexlab{}.
\newblock \showarticletitle{{Automated Software Winnowing}}.
\newblock \bibinfo{journal}{\emph{30th ACM Symposium on Applied Computing
  (SAC)}} (\bibinfo{year}{2015}).
\newblock
\urldef\tempurl%
\url{https://doi.org/10.1145/2695664.2695751}
\showURL{%
\tempurl}


\bibitem[Mishra and Polychronakis(2020)]%
        {Saffire}
\bibfield{author}{\bibinfo{person}{Shachee Mishra} {and}
  \bibinfo{person}{Michalis Polychronakis}.} \bibinfo{year}{2020}\natexlab{}.
\newblock \showarticletitle{Saffire: Context-sensitive Function Specialization
  against Code Reuse Attacks}. In \bibinfo{booktitle}{\emph{2020 IEEE European
  Symposium on Security and Privacy (EuroS\&P)}}. \bibinfo{pages}{17--33}.
\newblock
\urldef\tempurl%
\url{https://doi.org/10.1109/EuroSP48549.2020.00010}
\showDOI{\tempurl}


\bibitem[Mohagheghi et~al\mbox{.}(2004)]%
        {mohagheghi2004empirical}
\bibfield{author}{\bibinfo{person}{Parastoo Mohagheghi},
  \bibinfo{person}{Reidar Conradi}, \bibinfo{person}{Ole~M Killi}, {and}
  \bibinfo{person}{Henrik Schwarz}.} \bibinfo{year}{2004}\natexlab{}.
\newblock \showarticletitle{An empirical study of software reuse vs.
  defect-density and stability}. In \bibinfo{booktitle}{\emph{Proceedings. 26th
  International Conference on Software Engineering}}. IEEE,
  \bibinfo{pages}{282--291}.
\newblock


\bibitem[Navas and Gehani(2023)]%
        {OCCAM}
\bibfield{author}{\bibinfo{person}{Jorge~A. Navas} {and}
  \bibinfo{person}{Ashish Gehani}.} \bibinfo{year}{2023}\natexlab{}.
\newblock \showarticletitle{OCCAM-v2: Combining Static and Dynamic Analysis for
  Effective and Efficient Whole-Program Specialization}.
\newblock \bibinfo{journal}{\emph{Commun. ACM}} \bibinfo{volume}{66},
  \bibinfo{number}{4} (\bibinfo{date}{mar} \bibinfo{year}{2023}),
  \bibinfo{pages}{40–47}.
\newblock
\showISSN{0001-0782}
\urldef\tempurl%
\url{https://doi.org/10.1145/3583112}
\showDOI{\tempurl}


\bibitem[Pashakhanloo et~al\mbox{.}(2022)]%
        {PacJam}
\bibfield{author}{\bibinfo{person}{Pardis Pashakhanloo},
  \bibinfo{person}{Aravind Machiry}, \bibinfo{person}{Hyonyoung Choi},
  \bibinfo{person}{Anthony Canino}, \bibinfo{person}{Kihong Heo},
  \bibinfo{person}{Insup Lee}, {and} \bibinfo{person}{Mayur Naik}.}
  \bibinfo{year}{2022}\natexlab{}.
\newblock \showarticletitle{PacJam: Securing Dependencies Continuously via
  Package-Oriented Debloating}. In \bibinfo{booktitle}{\emph{ACM Asia
  Conference on Computer and Communications Security}} (Nagasaki, Japan)
  \emph{(\bibinfo{series}{ASIA CCS '22})}. \bibinfo{publisher}{Association for
  Computing Machinery}, \bibinfo{address}{New York, NY, USA},
  \bibinfo{pages}{903–916}.
\newblock
\showISBNx{9781450391405}
\urldef\tempurl%
\url{https://doi.org/10.1145/3488932.3524054}
\showDOI{\tempurl}


\bibitem[Porter et~al\mbox{.}(2023)]%
        {Decker}
\bibfield{author}{\bibinfo{person}{Chris Porter}, \bibinfo{person}{Sharjeel
  Khan}, {and} \bibinfo{person}{Santosh Pande}.}
  \bibinfo{year}{2023}\natexlab{}.
\newblock \showarticletitle{Decker: Attack Surface Reduction via On-Demand Code
  Mapping}. In \bibinfo{booktitle}{\emph{28th ACM International Conference on
  Architectural Support for Programming Languages and Operating Systems, Volume
  2}} (Vancouver, BC, Canada) \emph{(\bibinfo{series}{ASPLOS 2023})}.
  \bibinfo{publisher}{Association for Computing Machinery},
  \bibinfo{address}{New York, NY, USA}, \bibinfo{pages}{192–206}.
\newblock
\showISBNx{9781450399166}
\urldef\tempurl%
\url{https://doi.org/10.1145/3575693.3575734}
\showDOI{\tempurl}


\bibitem[Porter et~al\mbox{.}(2020)]%
        {BlankIt}
\bibfield{author}{\bibinfo{person}{Chris Porter}, \bibinfo{person}{Girish
  Mururu}, \bibinfo{person}{Prithayan Barua}, {and} \bibinfo{person}{Santosh
  Pande}.} \bibinfo{year}{2020}\natexlab{}.
\newblock \showarticletitle{BlankIt Library Debloating: Getting What You Want
  Instead of Cutting What You Don’t}. In \bibinfo{booktitle}{\emph{41st ACM
  SIGPLAN Conference on Programming Language Design and Implementation}}
  (London, UK) \emph{(\bibinfo{series}{PLDI 2020})}.
  \bibinfo{publisher}{Association for Computing Machinery},
  \bibinfo{address}{New York, NY, USA}, \bibinfo{pages}{164–180}.
\newblock
\showISBNx{9781450376136}
\urldef\tempurl%
\url{https://doi.org/10.1145/3385412.3386017}
\showDOI{\tempurl}


\bibitem[Qian et~al\mbox{.}(2019)]%
        {RAZOR}
\bibfield{author}{\bibinfo{person}{Chenxiong Qian}, \bibinfo{person}{Hong Hu},
  \bibinfo{person}{Mansour Alharthi}, \bibinfo{person}{Pak~Ho Chung},
  \bibinfo{person}{Taesoo Kim}, {and} \bibinfo{person}{Wenke Lee}.}
  \bibinfo{year}{2019}\natexlab{}.
\newblock \showarticletitle{{RAZOR}: A Framework for Post-deployment Software
  Debloating}. In \bibinfo{booktitle}{\emph{28th USENIX Security Symposium
  (USENIX Security 19)}}. \bibinfo{publisher}{USENIX Association},
  \bibinfo{address}{Santa Clara, CA}, \bibinfo{pages}{1733--1750}.
\newblock
\showISBNx{978-1-939133-06-9}


\bibitem[Qian et~al\mbox{.}(2020)]%
        {Slimium}
\bibfield{author}{\bibinfo{person}{Chenxiong Qian}, \bibinfo{person}{Hyungjoon
  Koo}, \bibinfo{person}{ChangSeok Oh}, \bibinfo{person}{Taesoo Kim}, {and}
  \bibinfo{person}{Wenke Lee}.} \bibinfo{year}{2020}\natexlab{}.
\newblock \showarticletitle{Slimium: Debloating the Chromium Browser with
  Feature Subsetting}. In \bibinfo{booktitle}{\emph{ACM SIGSAC Conference on
  Computer and Communications Security}} (Virtual Event, USA)
  \emph{(\bibinfo{series}{CCS '20})}. \bibinfo{publisher}{Association for
  Computing Machinery}, \bibinfo{address}{New York, NY, USA},
  \bibinfo{pages}{461–476}.
\newblock
\showISBNx{9781450370899}
\urldef\tempurl%
\url{https://doi.org/10.1145/3372297.3417866}
\showDOI{\tempurl}


\bibitem[Quach et~al\mbox{.}(2018)]%
        {Piece-Wise}
\bibfield{author}{\bibinfo{person}{Anh Quach}, \bibinfo{person}{Aravind
  Prakash}, {and} \bibinfo{person}{Lok Yan}.} \bibinfo{year}{2018}\natexlab{}.
\newblock \showarticletitle{Debloating Software through {Piece-Wise}
  Compilation and Loading}. In \bibinfo{booktitle}{\emph{27th USENIX Security
  Symposium (USENIX Security 18)}}. \bibinfo{publisher}{USENIX Association},
  \bibinfo{address}{Baltimore, MD}, \bibinfo{pages}{869--886}.
\newblock
\showISBNx{978-1-939133-04-5}


\bibitem[Rastogi et~al\mbox{.}(2017)]%
        {Cimplifier}
\bibfield{author}{\bibinfo{person}{Vaibhav Rastogi}, \bibinfo{person}{Drew
  Davidson}, \bibinfo{person}{Lorenzo De~Carli}, \bibinfo{person}{Somesh Jha},
  {and} \bibinfo{person}{Patrick McDaniel}.} \bibinfo{year}{2017}\natexlab{}.
\newblock \showarticletitle{Cimplifier: Automatically Debloating Containers}.
  In \bibinfo{booktitle}{\emph{11th Joint Meeting on Foundations of Software
  Engineering}} (Paderborn, Germany) \emph{(\bibinfo{series}{ESEC/FSE 2017})}.
  \bibinfo{publisher}{Association for Computing Machinery},
  \bibinfo{address}{New York, NY, USA}, \bibinfo{pages}{476–486}.
\newblock
\showISBNx{9781450351058}
\urldef\tempurl%
\url{https://doi.org/10.1145/3106237.3106271}
\showDOI{\tempurl}


\bibitem[Redini et~al\mbox{.}(2019)]%
        {BINTRIMMER}
\bibfield{author}{\bibinfo{person}{Nilo Redini}, \bibinfo{person}{Ruoyu Wang},
  \bibinfo{person}{Aravind Machiry}, \bibinfo{person}{Yan Shoshitaishvili},
  \bibinfo{person}{Giovanni Vigna}, {and} \bibinfo{person}{Christopher
  Kruegel}.} \bibinfo{year}{2019}\natexlab{}.
\newblock \showarticletitle{Bintrimmer: Towards static binary debloating
  through abstract interpretation}. In \bibinfo{booktitle}{\emph{Detection of
  Intrusions and Malware, and Vulnerability Assessment: 16th International
  Conference, DIMVA 2019, Gothenburg, Sweden, June 19--20, 2019, Proceedings
  16}}. Springer, \bibinfo{pages}{482--501}.
\newblock


\bibitem[Regehr et~al\mbox{.}(2012)]%
        {C-Reduce}
\bibfield{author}{\bibinfo{person}{John Regehr}, \bibinfo{person}{Yang Chen},
  \bibinfo{person}{Pascal Cuoq}, \bibinfo{person}{Eric Eide},
  \bibinfo{person}{Chucky Ellison}, {and} \bibinfo{person}{Xuejun Yang}.}
  \bibinfo{year}{2012}\natexlab{}.
\newblock \showarticletitle{Test-Case Reduction for C Compiler Bugs}. In
  \bibinfo{booktitle}{\emph{33rd ACM SIGPLAN Conference on Programming Language
  Design and Implementation}} (Beijing, China) \emph{(\bibinfo{series}{PLDI
  '12})}. \bibinfo{publisher}{Association for Computing Machinery},
  \bibinfo{address}{New York, NY, USA}, \bibinfo{pages}{335–346}.
\newblock
\showISBNx{9781450312059}
\urldef\tempurl%
\url{https://doi.org/10.1145/2254064.2254104}
\showDOI{\tempurl}


\bibitem[Shankar and Gehani(2012)]%
        {Shankar12}
\bibfield{author}{\bibinfo{person}{Natarajan Shankar} {and}
  \bibinfo{person}{Ashish Gehani}.} \bibinfo{year}{2012}\natexlab{}.
\newblock \showarticletitle{{Static Previrtualization}}.
\newblock \bibinfo{journal}{\emph{12th High Confidence Software and Systems
  Conference (HCSS)}} (\bibinfo{year}{2012}).
\newblock


\bibitem[Snow et~al\mbox{.}(2013)]%
        {snow2013just}
\bibfield{author}{\bibinfo{person}{Kevin~Z Snow}, \bibinfo{person}{Fabian
  Monrose}, \bibinfo{person}{Lucas Davi}, \bibinfo{person}{Alexandra
  Dmitrienko}, \bibinfo{person}{Christopher Liebchen}, {and}
  \bibinfo{person}{Ahmad-Reza Sadeghi}.} \bibinfo{year}{2013}\natexlab{}.
\newblock \showarticletitle{Just-in-time code reuse: On the effectiveness of
  fine-grained address space layout randomization}. In
  \bibinfo{booktitle}{\emph{2013 IEEE symposium on security and privacy}}.
  IEEE, \bibinfo{pages}{574--588}.
\newblock


\bibitem[Soto-Valero et~al\mbox{.}(2023)]%
        {JDBL}
\bibfield{author}{\bibinfo{person}{C\'{e}sar Soto-Valero},
  \bibinfo{person}{Thomas Durieux}, \bibinfo{person}{Nicolas Harrand}, {and}
  \bibinfo{person}{Benoit Baudry}.} \bibinfo{year}{2023}\natexlab{}.
\newblock \showarticletitle{Coverage-Based Debloating for Java Bytecode}.
\newblock \bibinfo{journal}{\emph{ACM Trans. Softw. Eng. Methodol.}}
  \bibinfo{volume}{32}, \bibinfo{number}{2}, Article \bibinfo{articleno}{38}
  (\bibinfo{date}{apr} \bibinfo{year}{2023}), \bibinfo{numpages}{34}~pages.
\newblock
\showISSN{1049-331X}
\urldef\tempurl%
\url{https://doi.org/10.1145/3546948}
\showDOI{\tempurl}


\bibitem[Soto-Valero et~al\mbox{.}(2021)]%
        {DEPCLEAN}
\bibfield{author}{\bibinfo{person}{C{\'e}sar Soto-Valero},
  \bibinfo{person}{Nicolas Harrand}, \bibinfo{person}{Martin Monperrus}, {and}
  \bibinfo{person}{Benoit Baudry}.} \bibinfo{year}{2021}\natexlab{}.
\newblock \showarticletitle{A comprehensive study of bloated dependencies in
  the maven ecosystem}.
\newblock \bibinfo{journal}{\emph{Empirical Software Engineering}}
  \bibinfo{volume}{26}, \bibinfo{number}{3} (\bibinfo{year}{2021}),
  \bibinfo{pages}{45}.
\newblock


\bibitem[Sun et~al\mbox{.}(2018)]%
        {Perses}
\bibfield{author}{\bibinfo{person}{Chengnian Sun}, \bibinfo{person}{Yuanbo Li},
  \bibinfo{person}{Qirun Zhang}, \bibinfo{person}{Tianxiao Gu}, {and}
  \bibinfo{person}{Zhendong Su}.} \bibinfo{year}{2018}\natexlab{}.
\newblock \showarticletitle{Perses: Syntax-Guided Program Reduction}. In
  \bibinfo{booktitle}{\emph{2018 IEEE/ACM 40th International Conference on
  Software Engineering (ICSE)}}. \bibinfo{pages}{361--371}.
\newblock
\urldef\tempurl%
\url{https://doi.org/10.1145/3180155.3180236}
\showDOI{\tempurl}


\bibitem[Tang et~al\mbox{.}(2022)]%
        {XDebloat}
\bibfield{author}{\bibinfo{person}{Yutian Tang}, \bibinfo{person}{Hao Zhou},
  \bibinfo{person}{Xiapu Luo}, \bibinfo{person}{Ting Chen},
  \bibinfo{person}{Haoyu Wang}, \bibinfo{person}{Zhou Xu}, {and}
  \bibinfo{person}{Yan Cai}.} \bibinfo{year}{2022}\natexlab{}.
\newblock \showarticletitle{XDebloat: Towards Automated Feature-Oriented App
  Debloating}.
\newblock \bibinfo{journal}{\emph{IEEE Transactions on Software Engineering}}
  \bibinfo{volume}{48}, \bibinfo{number}{11} (\bibinfo{year}{2022}),
  \bibinfo{pages}{4501--4520}.
\newblock
\urldef\tempurl%
\url{https://doi.org/10.1109/TSE.2021.3120213}
\showDOI{\tempurl}


\bibitem[Wang et~al\mbox{.}(2023)]%
        {Picup}
\bibfield{author}{\bibinfo{person}{Xiaoke Wang}, \bibinfo{person}{Tao Hui},
  \bibinfo{person}{Lei Zhao}, {and} \bibinfo{person}{Yueqiang Cheng}.}
  \bibinfo{year}{2023}\natexlab{}.
\newblock \showarticletitle{Input-Driven Dynamic Program Debloating for
  Code-Reuse Attack Mitigation}. In \bibinfo{booktitle}{\emph{31st ACM Joint
  European Software Engineering Conference and Symposium on the Foundations of
  Software Engineering}} \emph{(\bibinfo{series}{ESEC/FSE 2023})}.
  \bibinfo{publisher}{Association for Computing Machinery},
  \bibinfo{address}{New York, NY, USA}, \bibinfo{pages}{934–946}.
\newblock
\showISBNx{9798400703270}
\urldef\tempurl%
\url{https://doi.org/10.1145/3611643.3616274}
\showDOI{\tempurl}


\bibitem[Williams et~al\mbox{.}(2021)]%
        {PRAT}
\bibfield{author}{\bibinfo{person}{Ryan Williams}, \bibinfo{person}{Tongwei
  Ren}, \bibinfo{person}{Lorenzo De~Carli}, \bibinfo{person}{Long Lu}, {and}
  \bibinfo{person}{Gillian Smith}.} \bibinfo{year}{2021}\natexlab{}.
\newblock \showarticletitle{Guided Feature Identification and Removal for
  Resource-Constrained Firmware}.
\newblock \bibinfo{journal}{\emph{ACM Trans. Softw. Eng. Methodol.}}
  \bibinfo{volume}{31}, \bibinfo{number}{2}, Article \bibinfo{articleno}{28}
  (\bibinfo{date}{dec} \bibinfo{year}{2021}), \bibinfo{numpages}{25}~pages.
\newblock
\showISSN{1049-331X}
\urldef\tempurl%
\url{https://doi.org/10.1145/3487568}
\showDOI{\tempurl}


\bibitem[Wu et~al\mbox{.}(2021)]%
        {LightBlue}
\bibfield{author}{\bibinfo{person}{Jianliang Wu}, \bibinfo{person}{Ruoyu Wu},
  \bibinfo{person}{Daniele Antonioli}, \bibinfo{person}{Mathias Payer},
  \bibinfo{person}{Nils~Ole Tippenhauer}, \bibinfo{person}{Dongyan Xu},
  \bibinfo{person}{Dave~(Jing) Tian}, {and} \bibinfo{person}{Antonio Bianchi}.}
  \bibinfo{year}{2021}\natexlab{}.
\newblock \showarticletitle{{LIGHTBLUE}: Automatic {Profile-Aware} Debloating
  of Bluetooth Stacks}. In \bibinfo{booktitle}{\emph{30th USENIX Security
  Symposium (USENIX Security 21)}}. \bibinfo{publisher}{USENIX Association},
  \bibinfo{pages}{339--356}.
\newblock
\showISBNx{978-1-939133-24-3}


\bibitem[Xin et~al\mbox{.}(2021)]%
        {DomGad}
\bibfield{author}{\bibinfo{person}{Qi Xin}, \bibinfo{person}{Myeongsoo Kim},
  \bibinfo{person}{Qirun Zhang}, {and} \bibinfo{person}{Alessandro Orso}.}
  \bibinfo{year}{2021}\natexlab{}.
\newblock \showarticletitle{Subdomain-Based Generality-Aware Debloating}. In
  \bibinfo{booktitle}{\emph{35th IEEE/ACM International Conference on Automated
  Software Engineering}} (Virtual Event, Australia) \emph{(\bibinfo{series}{ASE
  '20})}. \bibinfo{publisher}{Association for Computing Machinery},
  \bibinfo{address}{New York, NY, USA}, \bibinfo{pages}{224–236}.
\newblock
\showISBNx{9781450367684}
\urldef\tempurl%
\url{https://doi.org/10.1145/3324884.3416644}
\showDOI{\tempurl}


\bibitem[Xin et~al\mbox{.}(2023)]%
        {Generality}
\bibfield{author}{\bibinfo{person}{Qi Xin}, \bibinfo{person}{Qirun Zhang},
  {and} \bibinfo{person}{Alessandro Orso}.} \bibinfo{year}{2023}\natexlab{}.
\newblock \showarticletitle{Studying and Understanding the Tradeoffs Between
  Generality and Reduction in Software Debloating}. In
  \bibinfo{booktitle}{\emph{37th IEEE/ACM International Conference on Automated
  Software Engineering}} (Rochester, MI, USA) \emph{(\bibinfo{series}{ASE
  '22})}. \bibinfo{publisher}{Association for Computing Machinery},
  \bibinfo{address}{New York, NY, USA}, Article \bibinfo{articleno}{99},
  \bibinfo{numpages}{13}~pages.
\newblock
\showISBNx{9781450394758}
\urldef\tempurl%
\url{https://doi.org/10.1145/3551349.3556970}
\showDOI{\tempurl}


\bibitem[Ye et~al\mbox{.}(2021)]%
        {JSLIM}
\bibfield{author}{\bibinfo{person}{Renjun Ye}, \bibinfo{person}{Liang Liu},
  \bibinfo{person}{Simin Hu}, \bibinfo{person}{Fangzhou Zhu},
  \bibinfo{person}{Jingxiu Yang}, {and} \bibinfo{person}{Feng Wang}.}
  \bibinfo{year}{2021}\natexlab{}.
\newblock \showarticletitle{JSLIM: Reducing the known vulnerabilities of
  Javascript application by debloating}. In
  \bibinfo{booktitle}{\emph{International Symposium on Emerging Information
  Security and Applications}}. Springer, \bibinfo{pages}{128--143}.
\newblock


\bibitem[Zhang et~al\mbox{.}(2024)]%
        {mmlb}
\bibfield{author}{\bibinfo{person}{Huaifeng Zhang}, \bibinfo{person}{Mohannad
  Alhanahnah}, \bibinfo{person}{Fahmi~Abdulqadir Ahmed}, \bibinfo{person}{Dyako
  Fatih}, \bibinfo{person}{Philipp Leitner}, {and} \bibinfo{person}{Ahmed
  Ali-Eldin}.} \bibinfo{year}{2024}\natexlab{}.
\newblock \showarticletitle{Machine Learning Systems are Bloated and
  Vulnerable}.
\newblock \bibinfo{journal}{\emph{Proc. ACM Meas. Anal. Comput. Syst.}}
  \bibinfo{volume}{8}, \bibinfo{number}{1}, Article \bibinfo{articleno}{6}
  (\bibinfo{date}{feb} \bibinfo{year}{2024}), \bibinfo{numpages}{30}~pages.
\newblock
\urldef\tempurl%
\url{https://doi.org/10.1145/3639032}
\showDOI{\tempurl}


\bibitem[Zhang et~al\mbox{.}(2023)]%
        {blafs}
\bibfield{author}{\bibinfo{person}{Huaifeng Zhang}, \bibinfo{person}{Mohannad
  Alhanahnah}, {and} \bibinfo{person}{Ahmed Ali-Eldin}.}
  \bibinfo{year}{2023}\natexlab{}.
\newblock \showarticletitle{BLAFS: A Bloat Aware File System}.
\newblock  (\bibinfo{year}{2023}).
\newblock
\showeprint[arxiv]{2305.04641}


\bibitem[Zhang et~al\mbox{.}(2022)]%
        {mTrimmer}
\bibfield{author}{\bibinfo{person}{Haotian Zhang}, \bibinfo{person}{Mengfei
  Ren}, \bibinfo{person}{Yu Lei}, {and} \bibinfo{person}{Jiang Ming}.}
  \bibinfo{year}{2022}\natexlab{}.
\newblock \showarticletitle{One Size Does Not Fit All: Security Hardening of
  MIPS Embedded Systems via Static Binary Debloating for Shared Libraries}. In
  \bibinfo{booktitle}{\emph{27th ACM International Conference on Architectural
  Support for Programming Languages and Operating Systems}} (Lausanne,
  Switzerland) \emph{(\bibinfo{series}{ASPLOS '22})}.
  \bibinfo{publisher}{Association for Computing Machinery},
  \bibinfo{address}{New York, NY, USA}, \bibinfo{pages}{255–270}.
\newblock
\showISBNx{9781450392051}
\urldef\tempurl%
\url{https://doi.org/10.1145/3503222.3507768}
\showDOI{\tempurl}


\end{thebibliography}
\end{document}